\documentclass[aps, prd, twocolumn, nofootinbib]{revtex4-2}

\usepackage{amsmath}
\usepackage{amssymb}
\usepackage{amsfonts}
\usepackage{wasysym}
\usepackage{graphicx}
\usepackage{comment}
\usepackage{tensor}
\usepackage[svgnames]{xcolor}

\usepackage[colorlinks = true, linkcolor = DarkBlue, citecolor = DarkBlue, urlcolor = DarkBlue, bookmarks = false]{hyperref}

\def\filetype{pdf}
\def\path{}

\begin{document}



\title{Are Einstein-Dirac-Maxwell wormholes traversable?}
\author{Ben Kain}
\affiliation{Department of Physics, College of the Holy Cross, Worcester, Massachusetts 01610, USA}

\begin{abstract}
\noindent Einstein-Dirac-Maxwell wormholes are asymptotically flat static wormhole solutions in general relativity that do not make use of exotic matter.  The asymmetric static solutions are smooth, are regular everywhere, and violate the null energy condition, which suggests that they are traversable.  To determine if in fact they are traversable, we numerically evolve the static solutions forward in time.  In all cases considered, our simulations indicate that black holes form that are connected by the wormhole.  Although there exist null geodesics that travel through the wormhole, we find that they are trapped inside a black hole and are unable to travel arbitrarily far away.  We conclude that Einstein-Dirac-Maxwell wormholes are not traversable.
\end{abstract} 

\maketitle


\section{Introduction}

Traversable wormholes are hypothetical gravitational systems that offer the possibility of quickly traveling between separated points in spacetime or even between different universes \cite{VisserBook, LoboBook}.  Asymptotically flat wormhole solutions have typically required exotic matter or modifications to general relativity.  For example, wormholes in asymptotically flat general relativity have been found with ghost matter, which has negative kinetic energy \cite{Ellis:1973yv, Bronnikov:1973fh, Armendariz-Picon:2002gjc}.  Modifications to general relativity that have led to wormholes or wormholes in anti--de Sitter space include Refs.~\cite{Bronnikov:2002rn, McFadden:2004ni, Kanti:2011jz, Kanti:2011yv, Gao:2016bin,  Maldacena:2018gjk, Maldacena:2020sxe}.

Recently, Bl\'azquez-Salcedo et al.\ discovered asymptotically flat static wormhole solutions in general relativity without making use of exotic matter \cite{Blazquez-Salcedo:2020czn, Blazquez-Salcedo:2021udn}.  They studied the Einstein-Dirac-Maxwell (EDM) system \cite{Finster:1998ws, Finster:1998ux, Finster:1998ju, Kain:2023jgu}, in which the charged Dirac equation is minimally coupled to general relativity, and found wormhole solutions that are symmetric across the wormhole throat.  The EDM system describes fermions with first quantized wave functions and treats gravity and the electromagnetic field classically.  Symmetric EDM wormholes, however, have concerning properties \cite{Danielson:2021aor}, which are absent in the asymmetric solutions found by Konoplya and Zhidenko \cite{Konoplya:2021hsm}.  In this work, we focus exclusively on asymmetric EDM wormholes.  For additional work on EDM wormholes, see \cite{Bolokhov:2021fil, Stuchlik:2021guq, Churilova:2021tgn, Wang:2022aze}.  

The authors of \cite{Blazquez-Salcedo:2020czn, Konoplya:2021hsm} assumed that EDM wormholes are traversable.  They presumably did so because their static solutions violate the null-energy condition \cite{Morris:1988cz} and are regular everywhere.  To determine if in fact EDM wormholes are traversable, we construct a time dependent EDM model.  We then use the asymmetric solutions as initial data and numerically evolve them forward in time.

Our simulations indicate that black holes form that are connected by the wormhole.  We find that null geodesics that can travel across the wormhole get trapped inside a black hole.  We see this behavior in all simulations we have performed.  For a wormhole to be traversable, one must be able to travel arbitrarily far on the opposite side.  While we cannot rule out that there may exist an asymmetric EDM wormhole that exhibits different behavior, our simulations lead us to conclude that EDM wormholes are not traversable.

In the next section, we present the Lagrangian for our matter sector and give the time dependent equations we solve for our simulations.  In Sec.~\ref{sec:semi-classical}, we explain how fermions are described by quantum wave functions.  In Sec.~\ref{sec:static}, we review asymmetric static EDM wormhole solutions.  Our main results are presented in Sec.~\ref{sec:simulations}, where we use static solutions as initial data and then numerically evolve them forward in time.  We conclude in Sec.~\ref{sec:conclusion}.  We have placed a number of details in appendices, including the most general set of spherically symmetric equations, our numerical methods, and code tests.  Unless otherwise indicated, we use units such that $c = G =\hbar = 1$.


\section{Einstein-Dirac-Maxwell}
\label{sec:EDM}

EDM wormholes are solutions to the Dirac equation gauged under $U(1)$ minimally coupled to spherically symmetric general relativity.  The matter sector includes charged spin-1/2 fermions, $\psi_x$, where $x$ labels the fermion, and a $U(1)$ gauge field, $\mathcal{A}_\mu$, and has Lagrangian
\begin{equation} \label{Lagrangian}
\mathcal{L} = \sum_x \mathcal{L}_\psi^x + \mathcal{L}_\mathcal{A},
\end{equation}
where
\begin{equation} \label{Lagrangian 2}
\begin{split}
\mathcal{L}_\psi^x &= \frac{1}{2} \left[
\bar{\psi}_x \gamma^\mu D_\mu \psi_x
- (D_\mu \bar{\psi}_x) \gamma^\mu \psi_x
\right]
- \mu \bar{\psi}_x \psi_x
\\
\mathcal{L}_\mathcal{A} &= -\frac{1}{4} F_{\mu\nu} F^{\mu\nu}
\\
F_{\mu\nu} &= \partial_\mu \mathcal{A}_\nu - \partial_\nu \mathcal{A}_\mu.
\end{split}
\end{equation}
In Appendices \ref{app:metric} and \ref{app:equations}, we show that spherical symmetry requires more than one fermion.  We include two fermions, which we label as $x = \pm$, both with mass $\mu$ and charge $e$.  Also in Appendices \ref{app:metric} and \ref{app:equations}, we give the definition of the covariant derivatives and the $\gamma$-matrices in Eq.\ (\ref{Lagrangian 2}) and present the complete set of equations for metric fields, the extrinsic curvature, and the equations of motion for the general spherically symmetric metric.  Different coordinate choices have different advantages in numerical simulations.  We choose to set the shift vector equal to zero, $\beta^r = 0$, which simplifies the equations.  In this section, we list the equations for this coordinate choice and we refer the reader to Appendices \ref{app:metric} and \ref{app:equations} for details.

In our time dependent model, we write the spherically symmetric metric as
\begin{equation}
ds^2 = -\alpha^2 dt^2 + A dr^2 + C \left( d\theta^2 + \sin^2\theta \, d\phi^2 \right),
\end{equation}
where $\alpha(t,r)$, $A(t,r)$, and $C(t,r)$ parametrize the metric.    Additionally we have $K\indices{^r_r}(t,r)$ and $K_T(t,r)$, which parametrize the extrinsic curvature.  In a wormhole geometry, $-\infty < r < \infty$.  We define
\begin{equation}
R(t,r) \equiv \sqrt{C(t,r)}
\end{equation}
as the areal radius so that the area of a two-sphere is $4\pi R^2$.  The minimum of the areal radius will occur at $r=0$ and we define
\begin{equation}
R_\text{th}(t) \equiv R(t,0)
\end{equation}
as the wormhole throat radius on a given time slice.

The Einstein field equations, $G\indices{^\mu_\nu} = 8\pi T\indices{^\mu_\nu}$, where $G\indices{^\mu_\nu}$ is the Einstein tensor and $T\indices{^\mu_\nu}$ is the energy-momentum tensor, lead to the evolution equations \cite{kain}
\begin{align} \label{evo eqs beta0} 
\partial_t A &=   - 2\alpha A K\indices{^r_r}
\notag \\
\partial_t C &= -\alpha C  K_T
\notag \\
\partial_t K\indices{^r_r} 
&=
\alpha \biggl[ \frac{(\partial_r C)^2}{4AC^2}
- \frac{1}{C} 
+ (K\indices{^r_r})^2
- \frac{1}{4}K_T^2 
\notag \\
&\qquad 
+ 4\pi (S +  \rho - 2 S\indices{^r_r} )
\biggr]
- \frac{\partial_r^2 \alpha}{A} + \frac{(\partial_r \alpha)(\partial_r A)}{2A^2}
\notag \\
\partial_t K_T
&= \alpha \left[ \frac{1}{C} 
- \frac{(\partial_r C)^2 }{4AC^2}
+ \frac{3}{4} (K_T)^2 
+ 8\pi S\indices{^r_r}
\right]
\notag \\
&\qquad
- \frac{(\partial_r \alpha)(\partial_r C)}{AC}
\end{align}
and to the Hamiltonian and momentum constraint equations,
\begin{equation} \label{con eqs beta0}
\begin{split}
\partial_r^2 C &=
A + \frac{(\partial_r A)(\partial_r C)}{2A} 
+ \frac{(\partial_r C)^2}{4C} 
\\
&\qquad
+ ACK_T \left( K\indices{^r_r} + \frac{1}{4} K_T \right)
- 8\pi AC \rho
\\
\partial_r K_T &=  
\frac{\partial_r C}{2C} \left(2 K\indices{^r_r} -  K_T \right)  -8\pi S_r.
\end{split}
\end{equation}
The definition of the matter functions $\rho$, $S\indices{^r_r}$, $S$, and $S_r$, which are functions of the energy-momentum tensor, are given in Eq.\ (\ref{matter functions defs}).

We minimally couple the matter sector to gravity via $\mathcal{L} \rightarrow \sqrt{-\det(g_{\mu\nu})} \, \mathcal{L}$, where $\det(g_{\mu\nu})$ is the determinant of the metric.  In a spherically symmetric system, $\psi_\pm = \psi_\pm(t,r,\theta,\phi)$, $\mathcal{A}_\mu = \mathcal{A}_\mu(t,r)$, and
\begin{equation}
\mathcal{A}_\mu = (\mathcal{A}_t, \mathcal{A}_r, 0,0).
\end{equation}
In our model, we describe fermions with the Dirac spinor ansatz
\begin{equation} \label{Dirac ansatz}
\psi_\pm = \frac{e^{\pm i\phi/2}}{2\sqrt{\pi} A^{1/4}(t,r)C^{1/2}(t,r)}
\begin{pmatrix}
\hphantom{i\pm} F(t,r) y_\pm(\theta) \\
\pm iF(t,r) y_\mp(\theta) \\
\hphantom{i\pm}G(t,r) y_\pm (\theta) \\
\mp iG(t,r) y_\mp(\theta)
\end{pmatrix},
\end{equation}
which is parametrized in terms of the complex fields $F$ and $G$, where
\begin{equation}
y_\pm(\theta) \equiv \sqrt{\frac{1 \mp \cos\theta}{2}} = 
\begin{array}{c}
\sin(\theta/2) \\ \cos(\theta/2)
\end{array}.
\end{equation}
The precise assumptions included in the ansatz are detailed in Appendix \ref{app:equations}.  Writing $F$ and $G$ in terms of their real and imaginary parts,
\begin{equation}
\begin{split}
F(t,r) &= F_1(t,r) + i F_2(t,r), 
\\
G(t,r) &= G_1(t,r) + i G_2(t,r),
\end{split}
\end{equation}
the equations of motion in the fermion sector are
\begin{widetext}
\begin{equation}
\begin{split}
\partial_t F_1 &= 
- \frac{\alpha G_1}{\sqrt{C}}
- \left[ F_2\left( e \mathcal{A}_t
- \mu \alpha\right) 
+G_2\frac{\alpha}{\sqrt{A}} e \mathcal{A}_r\right]
-  \frac{\alpha} {\sqrt{A}} \left( \partial_r G_1
+ \frac{\partial_r \alpha}{2\alpha} G_1 - \frac{\partial_r A}{4A} G_1
 \right)
\\
\partial_t F_2 &= 
- \frac{\alpha G_2}{\sqrt{C}}
+ \left[ F_1\left( e \mathcal{A}_t 
- \mu \alpha\right) 
+G_1\frac{\alpha}{\sqrt{A}} e \mathcal{A}_r\right]
-  \frac{\alpha} {\sqrt{A}} \left(\partial_r G_2
+ \frac{\partial_r \alpha}{2\alpha} G_2 - \frac{\partial_r A}{4A} G_2
 \right)
\\
\partial_t G_1 &= 
+ \frac{\alpha F_1}{\sqrt{C} }
- \left[G_2\left( e \mathcal{A}_t
+ \mu \alpha \right)
+ F_2  \frac{\alpha}{\sqrt{A}}  e \mathcal{A}_r
\right]
-  \frac{\alpha }{\sqrt{A} } \left(\partial_r F_1
+ \frac{\partial_r \alpha}{2\alpha}F_1 - \frac{\partial_r A}{4A} F_1
 \right)
\\
\partial_t G_2 &= 
+ \frac{\alpha F_2}{\sqrt{C} }
+ \left[G_1\left( e \mathcal{A}_t
+ \mu \alpha \right)
+ F_1  \frac{\alpha}{\sqrt{A}}  e \mathcal{A}_r
\right]
-  \frac{\alpha }{\sqrt{A} } \left(\partial_r F_2
+ \frac{\partial_r \alpha}{2\alpha}F_2 - \frac{\partial_r A}{4A} F_2
 \right).
\end{split}
\end{equation}
\end{widetext}

In the boson sector, it is convenient to choose a particular $U(1)$ gauge.  We choose to work in Lorenz gauge for our numerical simulations.  The equations of motion lead to
\begin{equation}
\begin{split}
\mathcal{A}_t &= \frac{\alpha}{\sqrt{A} C} \Omega
\\
\partial_t \mathcal{A}_r &= \frac{\alpha \sqrt{A}}{C} Y + \partial_r \mathcal{A}_t
\\
\partial_t \Omega &= \partial_r \left( \frac{\alpha C}{\sqrt{A}} \mathcal{A}_r  \right)
\\
\partial_t Y &= -\alpha \sqrt{A} C J^r,
\end{split}
\end{equation}
where $\Omega(t,r)$ and $Y(t,r)$ are auxiliary fields and where $J^r$ is the $r$-component of the conserved current,
\begin{equation} \label{Jt Jr}
J^r = - \frac{e}{\pi A C}
 \left(F_1 G_1 + F_2 G_2 \right) .
\end{equation}

Finally, the matter functions used in Eqs.\ (\ref{evo eqs beta0}) and (\ref{con eqs beta0}) contain contributions from both fermions and bosons,
\begin{equation}
\begin{split}
\rho &= \rho_\psi + \rho_\mathcal{A}
\\
S\indices{^r_r} &= (S_\psi)\indices{^r_r} + (S_\mathcal{A})\indices{^r_r}
\\
S &= S_\psi + S_\mathcal{A},
\\
S_r &= (S_\psi)_r + (S_\mathcal{A})_r,
\end{split}
\end{equation}
where the individual contributions are given in Eqs.\ (\ref{fermion matter functions}) and (\ref{boson matter functions}).


\section{Semiclassical theory}
\label{sec:semi-classical}


\subsection{Quantum wave functions}

Fermions obey the Pauli exclusion principle. By imposing the requirement that there be one fermion of each type,
\begin{equation} \label{N pm 1}
N_\pm = 1,
\end{equation}
we describe the Dirac fields with first quantized wave functions \cite{Finster:1998ws, Herdeiro:2017fhv, Blazquez-Salcedo:2020czn, Blazquez-Salcedo:2021udn}.  Just as in \cite{Blazquez-Salcedo:2020czn, Konoplya:2021hsm}, we ignore second quantization effects and treat gravity and the electromagnetic field classically.

To derive an equation for the number of fermions, $N_\pm$, first note that the system has a $U(1)$ gauge symmetry, which leads to a conserved current.  This conserved current is written in its typical form in Eq.\ (\ref{conserved current}) (we use the form in Eq.\ (\ref{Jt Jr}) for numerical solutions).  Our system also has global $U(1)$ symmetries for each fermion, corresponding to fermion number conservation.  The associated conserved currents, written in their typical form, are
\begin{equation}
j^\mu_\pm = i \bar{\psi}_\pm \gamma^\mu \psi_\pm,
\end{equation}
which are just the individual terms in (\ref{conserved current}) with the charge divided out, as expected.  The total number of each type of fermion on a time slice is given by the corresponding conserved charges, which are computed from
\begin{equation}
N_\pm = \int dr d\theta d\phi \, \sqrt{\gamma} \, n_\mu j^\mu_\pm,
\end{equation}
where $\gamma = A C^2 \sin^2\theta$ is the determinant of the spatial metric and $n_\mu$ follows from the timelike unit vector normal to a time slice given in (\ref{n up mu}).  Using the Dirac spinor ansatz in (\ref{Dirac ansatz}), we find the same answer for $\psi_+$ and $\psi_-$,
\begin{equation} \label{N pm}
N_\pm = \int_{-\infty}^\infty  dr \, \mathcal{N}, 
\qquad
\mathcal{N} = F_1^2 + F_2^2 + G_1^2 + G_2^2,
\end{equation}
where $\mathcal{N}$ is the number density.  In the next subsection, we explain how we impose $N_\pm = 1$ on Eq.\ (\ref{N pm}).


\subsection{Scaling}

In numerical work, it is often convenient to use scaled quantities.  We choose to scale variables by
\begin{equation}
R_0 \equiv R_\text{th}(0) = \sqrt{C(0,0)},
\end{equation}
which is the initial wormhole throat radius.  We define the following scaled quantities, which are indicated with an overbar:
\begin{align} \label{scaling}
r &\equiv R_0 \, \bar{r},& \qquad
t &\equiv R_0 \, \bar{t},
\notag
\\
e &\equiv (\sqrt{G}/R_0) \, \bar{e},& \qquad
\mu &\equiv \bar{\mu}/R_0,
\notag
\\
F_{1,2} &\equiv \overline{F}_{1,2} / \sqrt{G/R_0},& \qquad
G_{1,2} &\equiv \overline{G}_{1,2} / \sqrt{G/R_0}, \qquad
\notag
\\
\mathcal{A}_{t,r} &\equiv \bar{\mathcal{A}}_{t,r} / \sqrt{G},& 
C &\equiv R_0^2 \, \overline{C}.
\end{align}
For convenience, we have included the gravitational constant $G = \ell_P^2$, where $\ell_P$ is the Planck length.  Upon rewriting all equations in terms of the scaled quantities, $R_0$ cancels out.  $R_0$ therefore does not have to be specified and our code works for arbitrary values of $R_0$.

The total number of each type of fermion is given by Eq.\ (\ref{N pm}), which scales to
\begin{equation}
\overline{N}_\pm  = \int_{-\infty}^\infty d\bar{r} \, \overline{\mathcal{N}},
\qquad
\overline{\mathcal{N}} = \overline{F}_1^2 + \overline{F}_2^2 + \overline{G}_1^2 + \overline{G}_2^2,
\end{equation}
where
\begin{equation}
N_\pm \equiv (R_0 / \ell_P)^2 \, \overline{N}_\pm.
\end{equation}
Setting $N_\pm = 1$ gives
\begin{equation} \label{R0 eq}
R_0 = \frac{\ell_P}{\sqrt{\overline{N}_\pm}}.
\end{equation}
The importance and convenience of this result is as follows.  With a quantum wave function, where $N_\pm = 1$, the value of the scaled quantity ${\overline{N}}_\pm$, which is computed by our code, tells us the physical value $R_0$.  In practice, then, our code does not explicitly  normalize the wave function.  Instead, it solves the equations written in terms of scaled variables and the value computed for ${\overline{N}}_\pm$, and hence for $R_0$ through Eq.\ (\ref{R0 eq}), singles out the specific physical semiclassical theory under study.


\section{Static wormholes}
\label{sec:static}

In this section, we present solutions for static EDM wormholes.  By static, we mean that spacetime is time independent.  We will use these static solutions as initial data for our simulations, but they are also interesting in their own right.  We focus exclusively on asymmetric EDM wormholes \cite{Konoplya:2021hsm}.


\subsection{Equations}

Static solutions are found by dropping the time dependence for the metric fields and setting $K\indices{^r_r} = K_T = 0$ (in the context of the static solution being used for initial data, we are assuming the initial data are time symmetric \cite{BaumgarteBook}). The energy-momentum tensor must be time independent and diagonal.  In the fermion sector, this can be achieved by assuming
\begin{equation} \label{static F G}
F(t,r) = f(r) e^{-i\omega t}, \qquad
G(t,r) = i g(r) e^{-i\omega t},
\end{equation}
where $f(r)$ and $g(r)$ are real functions and $\omega$ is a real constant.  In the boson sector, we take $\mathcal{A}_\mu$ to be time independent.  We choose to work in radial gauge, where
\begin{equation}
\mathcal{A}_r = 0,
\end{equation} 
which is consistent with the static limit of Lorenz gauge.  Note that $\omega$ is gauge dependent and can be shifted in Eq.\ (\ref{static F G}) with a simple gauge transformation that does not take us out of radial gauge.

The complete set of equations under these assumptions is given in Appendix \ref{app:static}.  The next step is to make a coordinate choice.  For example, the authors of \cite{Konoplya:2021hsm} make the choice $C = R_0^2 [1 - (r/R_0)^2]^{-2}$, which compactifies the radial coordinate to $-R_0 < r < R_0$.  A variation of this choice, which is used in \cite{Blazquez-Salcedo:2020czn}, is $C = R_0^2 + r^2$, which leaves the radial direction uncompactified.  We have been able to find static solutions using both choices.  However, we choose to set
\begin{equation} \label{A1}
A(r) = 1,
\end{equation}
which we find to be a convenient choice when using static solutions as initial data.  The static metric is then
\begin{equation}
ds^2 = -\alpha^2(r) dt^2 + dr^2 + C(r) (d\theta^2 + \sin^2\theta d\phi^2).
\end{equation}

We now list the equations given in Appendix \ref{app:static}, but with our coordinate choice in Eq.\ (\ref{A1}).  The metric equations are 
\begin{equation} \label{static metric eqs A1}
\begin{split}
\partial_r^2 C &= 1 + \frac{(\partial_r C)^2}{4 C} -8\pi C \rho
\\
\partial_r^2\alpha &= - \frac{(\partial_r\alpha) (\partial_r C)}{C} + 4\pi \alpha  (\rho + S)
\\
0 &=\frac{(\partial_r\alpha) (\partial_r C)}{\alpha C} - \frac{1}{C} + \frac{(\partial_r C)^2}{4 C^2} - 8\pi S\indices{^r_r},
\end{split}
\end{equation}
the equations of motion are 
\begin{equation} \label{statc eom A1}
\begin{split}
\partial_r f &= 
- f \left( \frac{ \partial_r\alpha}{2\alpha} - \frac{1}{\sqrt{C}}\right)
- g \left( \mu + \frac{u}{\alpha} \right)
\\
\partial_r g &= 
- g \left( \frac{\partial_r \alpha}{2\alpha} + \frac{1}{\sqrt{C}}\right)
- f \left( \mu - \frac{u}{\alpha} \right)
\\
\partial_r^2 u &= - \partial_r u \left(\frac{\partial_rC}{C} - \frac{\partial_r\alpha}{\alpha}  \right)
+ e^2 \frac{\alpha }{2\pi C} ( f^2 + g^2),
\end{split}
\end{equation}
where 
\begin{equation}
u(r) \equiv e \mathcal{A}_t(r) + \omega,
\end{equation}
and the matter functions are 
\begin{equation}
\begin{split}
\rho &=  \frac{u(f^2 + g^2)}{2\pi \alpha C}
+ \frac{(\partial_r u)^2}{2e^2\alpha^2} 
\\
S\indices{^r_r} &= \frac{f \partial_rg - g \partial_r f}{2\pi C} 
- \frac{(\partial_r u)^2}{2e^2\alpha^2} 
\\
S\indices{^\theta_\theta} &= \frac{f g}{2\pi C^{3/2}}
+ \frac{(\partial_r u)^2}{2e^2\alpha^2},
\end{split}
\end{equation}
along with $S = S\indices{^r_r} + 2S\indices{^\theta_\theta}$, where 
\begin{equation}
f\partial_r g - g \partial_rf
= -  \frac{2fg}{\sqrt{C}}
-  \mu (f^2 - g^2)+ \frac{u}{\alpha} (f^2 + g^2),
\end{equation}
which follows from the first two equations in (\ref{statc eom A1}).  The matter sector is described by the three fields $f(r)$, $g(r)$, and $u(r)$.  Although our equations look different than those in \cite{Konoplya:2021hsm}, we explain how they are equivalent in Appendix \ref{app:comparison}.  For static solutions, Eq.\ (\ref{N pm}), which computes the total number of each type of fermion, reduces to
\begin{equation} \label{static N pm}
N_\pm = \int_{-\infty}^\infty  dr \mathcal{N}(r), \qquad \mathcal{N} = f^2 + g^2,
\end{equation}
where $\mathcal{N}$ is the number density.

We move now to the scaled variables in Eq.\ (\ref{scaling}), along with
\begin{equation}
f \equiv \bar{f} / \sqrt{G/R_0}, \qquad
g \equiv \bar{g} / \sqrt{G/R_0}, \qquad
u \equiv \bar{u}/R_0.
\end{equation}
We solve the static equations by numerically integrating the top two equations in (\ref{static metric eqs A1}) and the equations of motion in (\ref{statc eom A1}) outward from $\bar{r} = 0$.  This requires conditions at $\bar{r} = 0$.  We assume
\begin{equation}
\overline{C}^{\, \prime}(0) = 0, \qquad
\alpha'(0) = 0,
\end{equation}
where a prime denotes an $\bar{r}$-derivative.  These same assumptions were made in \cite{Konoplya:2021hsm} (the possibility of $\alpha'(0) \neq 0$ was considered in \cite{Wang:2022aze}).  We have also $\overline{C}(0) = 1$ (since we scaled variables by $R_0$).  Inspection of the equations shows that if we define
\begin{equation}
\widetilde{u}(r) \equiv \frac{\bar{u}(r)}{\alpha(0)}, \qquad \widetilde{\alpha}(r) \equiv \frac{\alpha(r)}{\alpha(0)}
\end{equation}
and then write the equations in terms of $\widetilde{u}$ and $\widetilde{\alpha}$, $\alpha(0)$ cancels out and we have $\widetilde{\alpha}(0) = 1$.  Lastly, by plugging Taylor expansions of the fields into the bottom equation in (\ref{static metric eqs A1}), we are able to derive
\begin{equation}
\widetilde{u}^{\prime \, 2} (0) = \frac{\bar{e}^2}{4\pi}
\left[
1 - 8f_0 g_0 + 4 u_0 (f_0^2 + g_0^2) - 4\bar{\mu}(f_0^2 - g_0^2)
\right],
\end{equation}
where
\begin{equation} \label{f0 g0 u0}
f_0 \equiv \bar{f}(0), \qquad
g_0 \equiv \bar{g}(0), \qquad
u_0 \equiv \widetilde{u}(0).
\end{equation}

The three quantities in Eq.\ (\ref{f0 g0 u0}), along with $\bar{\mu}$ and $\bar{e}$, are the quantities that must be specified to be able to integrate the static equations outward from $\bar{r} = 0$.  We choose to identify static solutions with $f_0$ and take $g_0$ and $u_0$ to be shooting parameters.  That is, we specify $f_0$, $\bar{\mu}$, and $\bar{e}$ and then tune $g_0$ and $u_0$ using the shooting method until the integrated result satisfies the outer boundary conditions.  The outer boundary conditions are that spacetime is asymptotically flat, i.e.\ that $\rho \rightarrow 0$ as $r\rightarrow \pm \infty$, which requires $f,g,u' \rightarrow 0$.  Once the outer boundary conditions are satisfied, we have found a static wormhole solution.


\subsection{Results}
\label{sec:static results}

\begin{table}
\begin{tabular}{r@{\hskip 5pt}|@{\hskip 5pt}c@{\hskip 5pt}|@{\hskip 5pt}c@{\hskip 5pt}|@{\hskip 5pt}c}
$f_0$\hspace{8pt} & $g_0$ & $u_0$ & $R_0/\ell_P$ \\
\hline\hline
$0.001$ & $0.000702$ & $-2.519$ & $376.6$  \\
$0.002$ & $0.001404$ & $-2.519$ & $188.3$  \\
$0.003$ & $0.002105$ & $-2.519$ & $125.5$  \\
$0.004$ & $0.002806$ & $-2.518$ & $94.12$  \\
$0.005$ & $0.003507$ & $-2.518$ & $75.28$  \\
\hline
$-0.001$ & $0.001257$ & $-1.519$ & $498.4$  \\
$-0.002$ & $0.002514$ & $-1.519$ & $249.2$  \\
$-0.003$ & $0.003771$ & $-1.519$ & $166.1$  \\
$-0.004$ & $0.005028$ & $-1.519$ & $124.6$  \\
$-0.005$ & $0.006285$ & $-1.519$ & $99.66$  \\
\end{tabular}
\caption{Results for static asymmetric EDM wormhole solutions with $\bar{\mu} = 0.2$ and $\bar{e}/\sqrt{4\pi} = 0.03$.  Various fields for these solutions are plotted in Fig.\ \ref{fig:static solutions}.}
\label{table:static solutions}
\end{table}

\begin{figure*}
\centering
\includegraphics[width=7in]{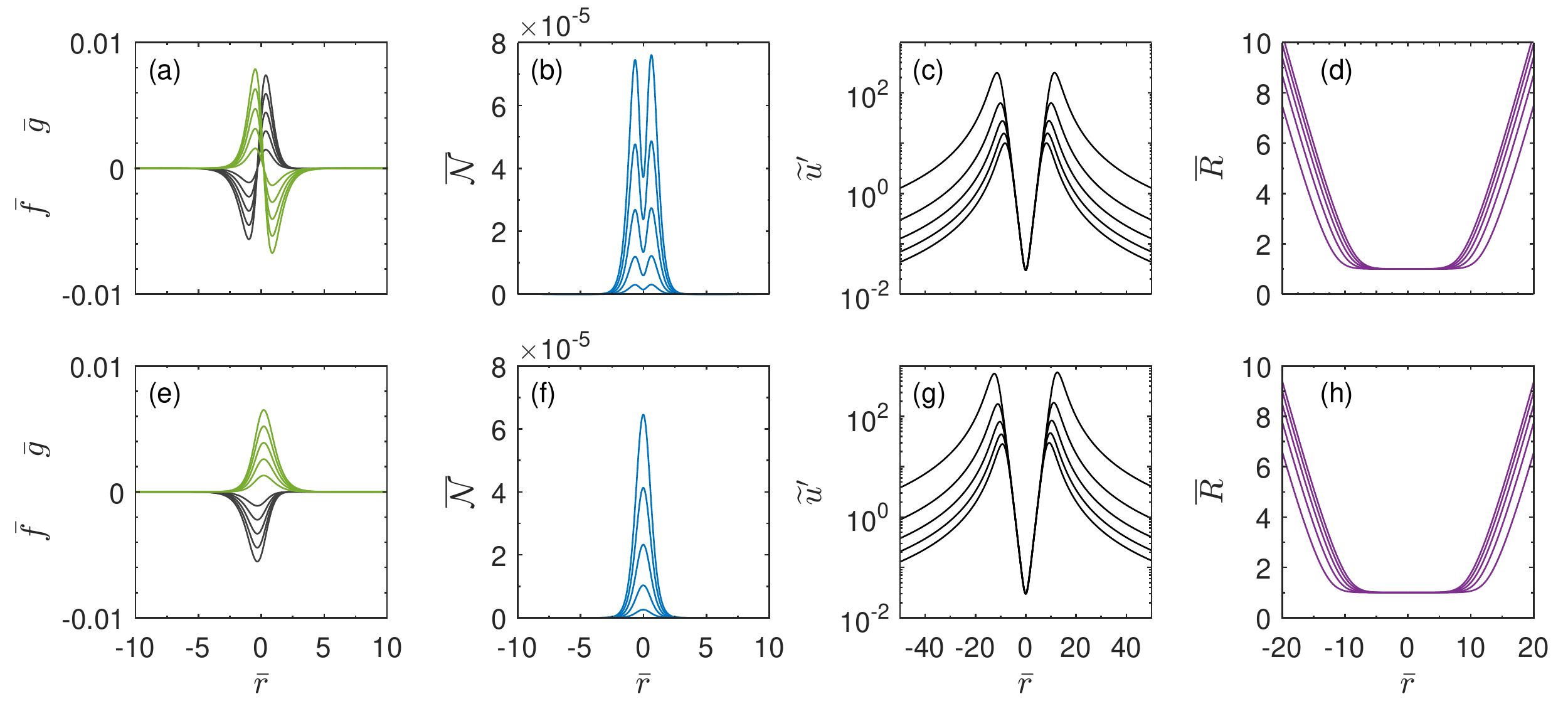}
\caption{Static asymmetric EDM wormhole solutions with $\bar{\mu} = 0.2$ and $\bar{e}/\sqrt{4\pi} = 0.03$.  These same solutions are listed in Table \ref{table:static solutions}.  The top row displays solutions with positive $f_0$ and the bottom row displays solutions with negative $f_0$.  As the magnitude of $f_0$ decreases, the magnitude of the peaks in (a), (e) and (b), (f) decrease, the peaks in (c), (g) increase, and the curves in (d), (h) widen.}
\label{fig:static solutions}
\end{figure*} 

\begin{figure}
\centering
\includegraphics[width=2.45in]{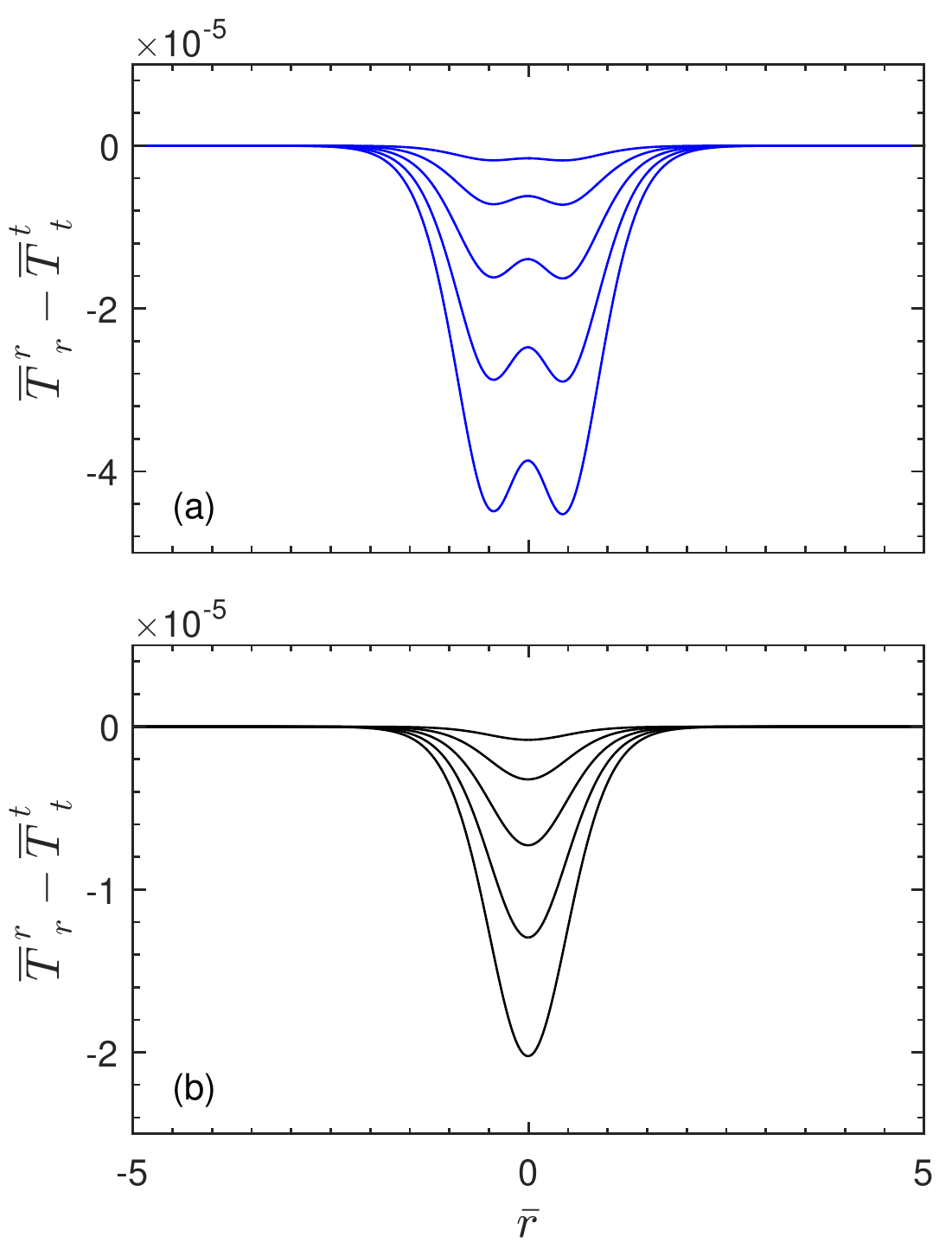}
\caption{The top rows of this plot and of Fig.\ \ref{fig:static solutions} display the same solutions and similarly for the bottom rows.  In all cases, we can see that the null energy condition is violated, which occurs if $\overline{T}\indices{^r_r} - \overline{T}\indices{^t_t} < 0$.}
\label{fig:static null}
\end{figure} 

The results for some typical static solutions are listed in Table \ref{table:static solutions} for $\bar{\mu} = 0.2$ and $\bar{e}/\sqrt{4\pi} = 0.03$ ($\bar{e}/\sqrt{4\pi}$ is the convention used in \cite{Konoplya:2021hsm} for charge, which is the reason why we include the factor of $\sqrt{4\pi}$).  We show in Fig.\ \ref{fig:static solutions} some of the fields for these solutions.  The top row of Fig.\ \ref{fig:static solutions} displays solutions with positive $f_0$ and the bottom row displays them for negative $f_0$.

In Figs.\ \ref{fig:static solutions}(a) and \ref{fig:static solutions}(e), the black curves show $\bar{f}(r)$ and the green curves show $\bar{g}(r)$.  We can see that the curves satisfy the outer boundary conditions.  Figures \ref{fig:static solutions}(b) and \ref{fig:static solutions}(f) plot the number density, $\overline{\mathcal{N}} = \bar{f}^2 + \bar{g}^2$.  Figures \ref{fig:static solutions}(c) and \ref{fig:static solutions}(g) plot $\widetilde{u}'$ and we can see that the curves are asymptotically heading to zero, consistent with the outer boundary conditions.  Finally, Figs.\ \ref{fig:static solutions}(d) and \ref{fig:static solutions}(h) plot the areal radius, $\overline{R}$, where our scaling convention fixes $\overline{R} = 1$ at $\bar{r}= 0$.

We mention two interesting points.  First, since we can interpret the number density as a particle distribution, the two peaks in Fig.\ \ref{fig:static solutions}(b) suggest an interpretation of two quasilocalized particles, with one particle on each side of the wormhole.  Second, the final column in Table \ref{table:static solutions} gives the physical value for $R_0$, as computed from Eq.\ (\ref{R0 eq}), which is the wormhole throat radius.  We can see that, for these solutions, the radius is roughly two orders of magnitude larger than the Planck length.

Morris and Thorne showed that traversable wormholes violate the null energy condition \cite{Morris:1988cz}.  To determine if our static solutions violate the null energy condition, we use radial null vectors $d^\mu = \lambda (1, \pm \alpha/\sqrt{A}, 0, 0)$, where $\lambda$ is an arbitrary constant.  The null energy condition is violated if $T_{\mu\nu} d^\mu d^\nu < 0$, or
\begin{equation} \label{null condition}
T\indices{^r_r} - T\indices{^t_t} < 0,
\end{equation}
where the energy-momentum tensor components for static solutions are given in Eqs.\ (\ref{static T psi}) and (\ref{static T A}).  Figure \ref{fig:static null} shows the results for Eq.\ (\ref{null condition}) for the same static solutions shown in Fig.\ \ref{fig:static solutions} and listed in Table \ref{table:static solutions}.  We can see that the static solutions violate the null energy condition.

In this section, we presented some typical static EDM wormhole solutions.  The solutions are regular everywhere and violate the null energy condition.  As a consequence, it seems plausible that EDM wormholes are traversable, as is assumed in Refs.\ \cite{Blazquez-Salcedo:2020czn, Konoplya:2021hsm}.  To determine if in fact EDM wormholes are traversable, in the next section we use these static solutions as initial data and numerically evolve them forward in time.  We then compute null geodesics and study how the geodesics travel through the wormhole.


\section{Wormhole simulations}
\label{sec:simulations}

\begin{figure*}
\centering
\includegraphics[width=7in]{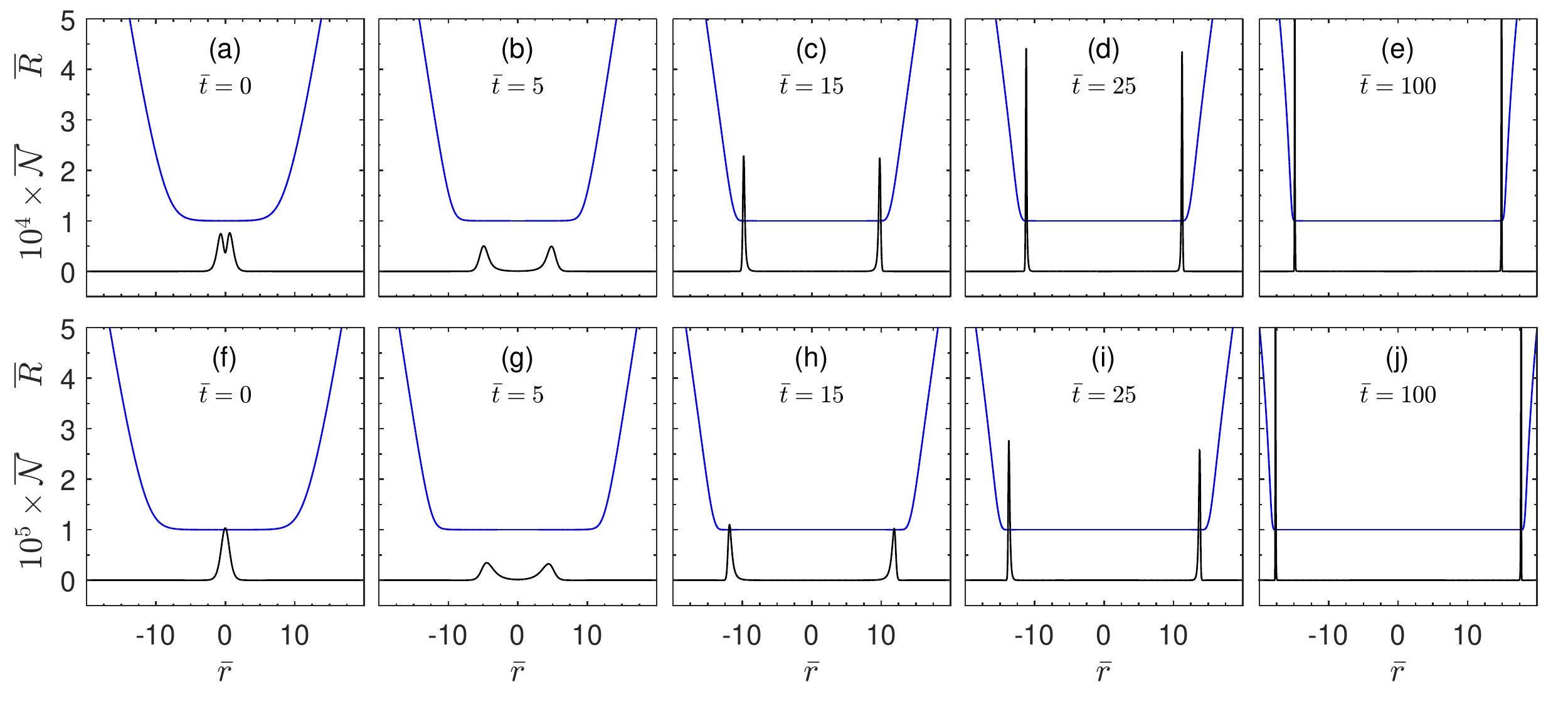}
\caption{The top row displays a numerical evolution of an EDM wormhole using the static solution from Sec.\ \ref{sec:static results} with $f_0 = 0.005$.  The bottom displays a different evolution using the static solution with $f_0 = -0.002$.  The blue curves plot the areal radius, $\overline{R}$, and the black curves plot the fermion number density, $\overline{\mathcal{N}}$.}
\label{fig:R fsqgsq}
\end{figure*} 

\begin{figure}
\centering
\includegraphics[width=2.45in]{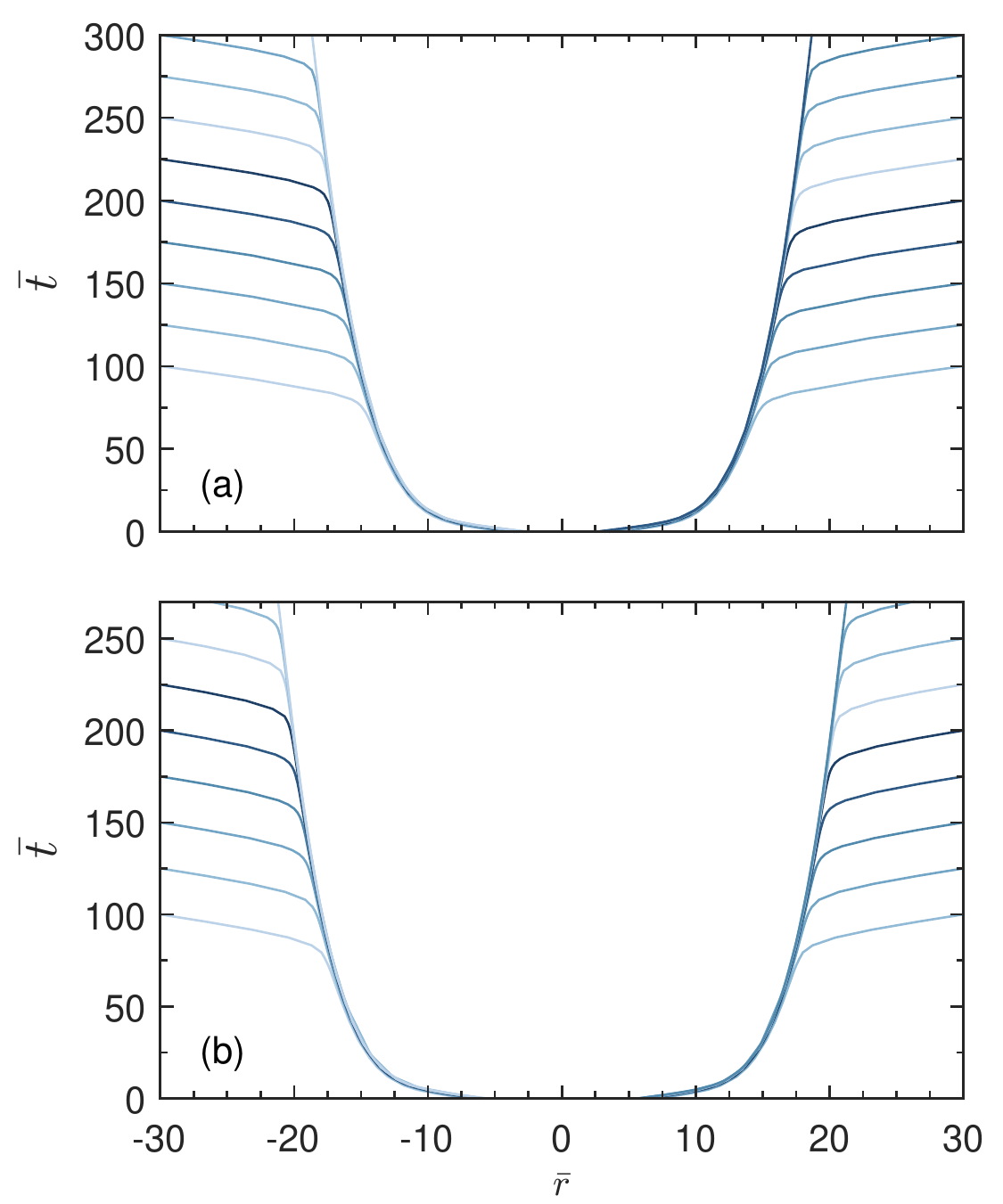}
\caption{(a) Null geodesics are shown for the same simulation shown in the top row of Fig.\ \ref{fig:R fsqgsq}.  The geodesics map out event horizons and indicate the presence of black holes.  (b) The same as (a) except for the simulation shown in the bottom row of Fig.\ \ref{fig:R fsqgsq}.}
\label{fig:event horizon}
\end{figure}

In this section, we simulate EDM wormholes using the static solutions presented in the previous section as initial data.  This requires numerically solving the equations presented in Sec.\ \ref{sec:EDM}.  Our numerical methods, including the equations we use to compute null geodesics and apparent horizons, are described in Appendix \ref{app:numerics}.  In Appendix \ref{app:code tests}, we present tests of our code and show that our code is second order convergent.

We show results for two simulations in Fig.\ \ref{fig:R fsqgsq}.  The top row presents a single simulation using as initial data the static solution from the previous section with $f_0 = 0.005$ and the bottom row shows a different simulation using the static solution with $f_0 = -0.002$.  The blue curves plot the areal radius, $\overline{R}$, and the black curves plot the number density, $\overline{\mathcal{N}}$.  As time increases, with respect to their radial coordinates, the spikes in the number density separate and the length of the wormhole throat increases.  The wormhole throat radius holds steady, neither expanding nor collapsing.

In general, numerical simulations cannot definitively prove that a black hole is present, since such a proof requires knowledge of the entire spacetime.  The most conclusive method for determining if a black hole is present in a simulation is computing null geodesics backwards in time \cite{Anninos:1994ay, Libson:1994dk, BaumgarteBook}.  This method is particularly conclusive in spherically symmetric spacetimes, since null rays can only travel in the radial direction.  This is done after the simulation is complete from the stored results.  If a black hole is present, the computed null geodesics can determine the location of the event horizon.  We show such null geodesics in Fig.\ \ref{fig:event horizon}.  Figure \ref{fig:event horizon}(a) is for the same simulation shown in the top row of Fig.\ \ref{fig:R fsqgsq} and similarly for the bottom rows.  The null geodesics clearly indicate the presence of event horizons on both sides of the wormhole.  The wormhole therefore connects two black holes.  Although this is the most conclusive method, we present other indicators for the presence of a black hole in what follows.

\begin{figure}
\centering
\includegraphics[width=3in]{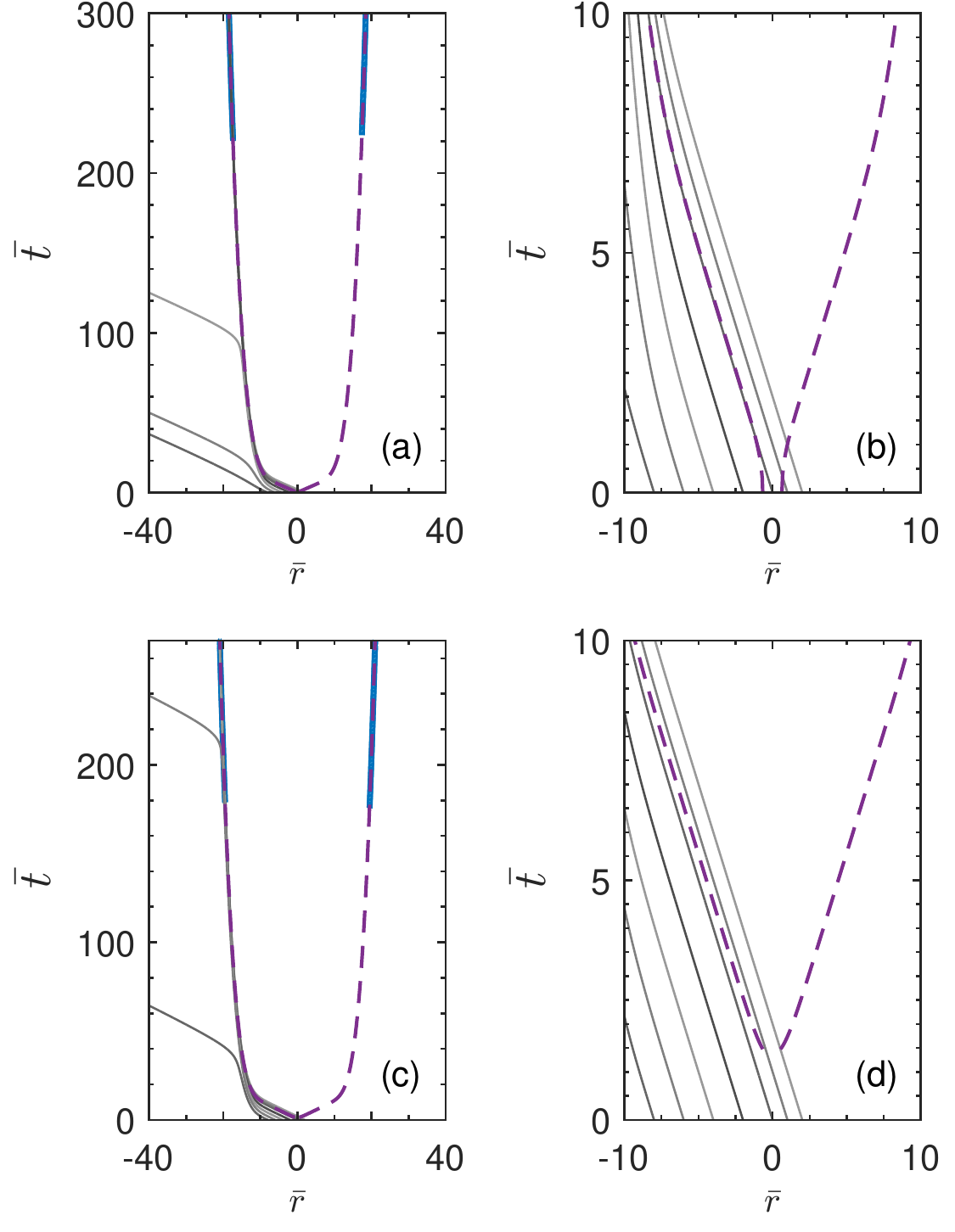}
\caption{The top row displays the same simulation as shown in the top row of Fig.\ \ref{fig:R fsqgsq}.  The purple dashed curves mark the position of the peaks of the spikes of the number density shown in Fig.\ \ref{fig:R fsqgsq}.  The blue curves plot apparent horizons.  The gray curves plot null geodesics.  (b) zooms into the early $\bar{t}$ region of (a).  The bottom row is analogous to the top row except it displays the same simulation shown in the bottom row of Fig.\ \ref{fig:R fsqgsq}.}
\label{fig:null ah}
\end{figure}

In Fig.\ \ref{fig:null ah}, we show another way to view the simulations shown in Fig.\ \ref{fig:R fsqgsq}.  The dashed purple lines in Figs.\ \ref{fig:null ah}(a) and \ref{fig:null ah}(b) track the peaks of the spikes of the number density in the top row of Fig.\ \ref{fig:R fsqgsq}.  The dashed purple lines can therefore be thought of as plotting the position of the two particles.  The bottom row plots the same thing as the top row, except for the simulation shown in the bottom row of Fig.\ \ref{fig:R fsqgsq}.  In Fig.\ \ref{fig:null ah}(d), we only begin plotting the dashed purple lines after a clear separation and the formation of two peaks in the number density.

The thick blue lines in Fig.\ \ref{fig:null ah}(a), which start at around $\bar{t} = 225$, plot apparent horizons.  Standard theorems in general relativity concerning apparent horizons do not necessarily apply to systems that violate the null energy condition (see the discussion in \cite{Gonzalez:2008xk}).  Nonetheless, numerical evidence suggests that the apparent horizon may still be a good indicator for the presence of a black hole \cite{Gonzalez:2008xk, kain}.

\begin{figure}
\centering
\includegraphics[width=2.45in]{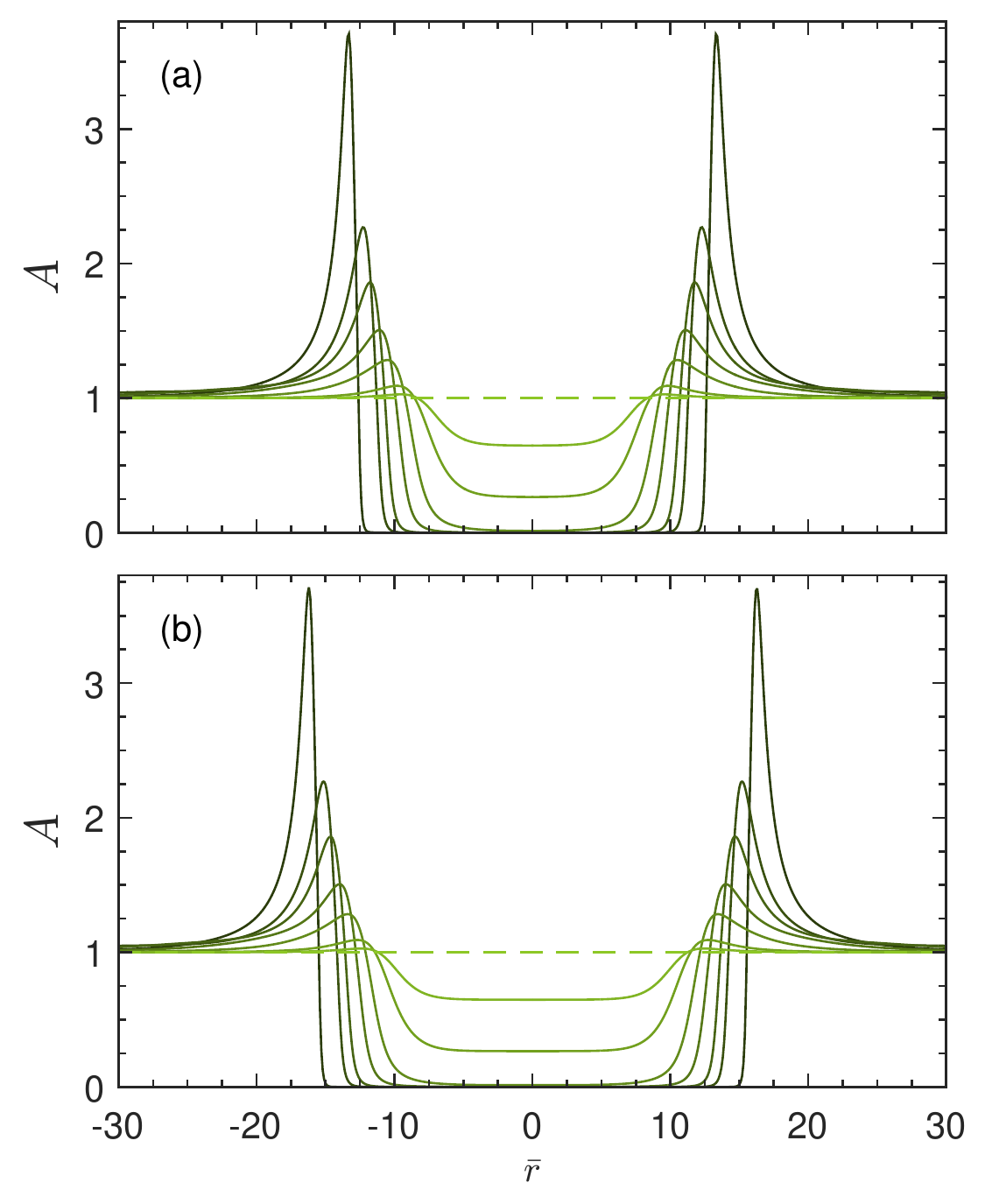}
\caption{(a) The metric field $A$ for the simulation shown in the top row of Fig.\ \ref{fig:R fsqgsq}.  The dashed line plots $A$ at $\bar{t} = 0$.  As time moves forward, $A$ drops in the middle toward zero and forms spikes at the edges.  (b) The same as (a), except for the simulation shown in the bottom row of  Fig.\ \ref{fig:R fsqgsq}.}
\label{fig:A}
\end{figure} 

We have computed various left-moving null geodesics and plotted them as the gray lines in Fig.\ \ref{fig:null ah}(a).  Figure \ref{fig:null ah}(b) zooms into the early $\bar{t}$ region of Fig.\ \ref{fig:null ah}(a), where we can see that two geodesics originate on the positive side of the wormhole, one originates from $\bar{r} = 0$, and four originate on the negative side.  Those that originate on the positive side can be seen to cross $\bar{r} = 0$, and hence to travel through the wormhole.  However, we find that any geodesic that originates on the positive side, and some that originate on the negative side, are unable to travel arbitrarily far on the negative side.  Indeed, we find that such geodesics get caught right up against the apparent horizon and are trapped inside the black hole.

\begin{figure}
\centering
\includegraphics[width=2.45in]{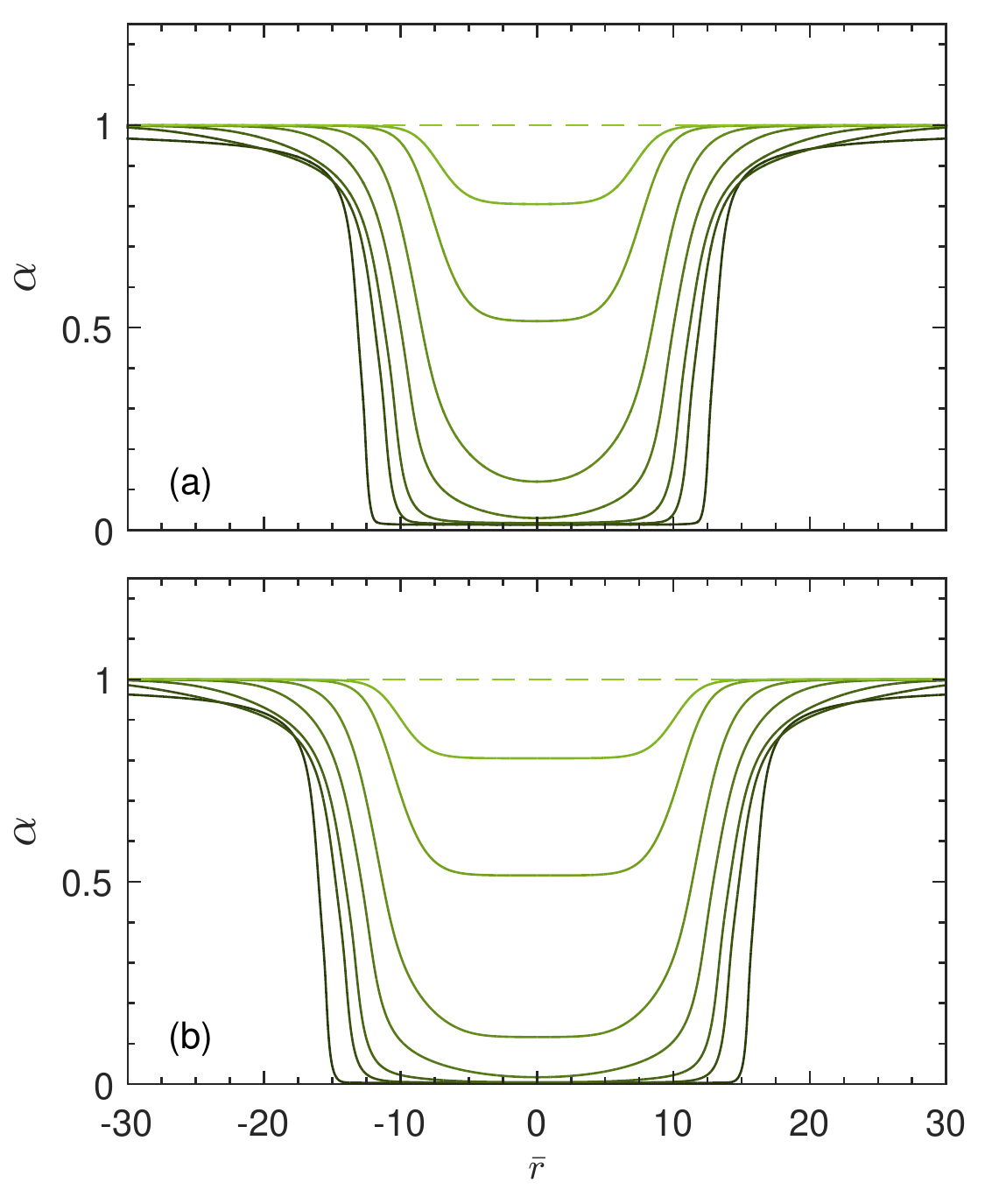}
\caption{(a) The metric field $\alpha$ for the simulation shown in the top row of Fig.\ \ref{fig:R fsqgsq}.  The dashed line plots $\alpha$ at $\bar{t} = 0$.  As time moves forward, $\alpha$ collapses in the middle.  (b) The same as (a), except for the simulation shown in the bottom row of Fig.\ \ref{fig:R fsqgsq}.}
\label{fig:alpha}
\end{figure}

We show the evolution of the metric field $A$ in Fig.\ \ref{fig:A} for the same simulations shown in Fig.\ \ref{fig:R fsqgsq}.  The horizontal dashed lines plot $A$ at $\bar{t} = 0$.  As time increases, $A$ collapses toward zero in the middle and forms spikes at the edges.  Since $A$ gives a measure of the physical distance between radial coordinates, that $A$ is collapsing means that the radial direction is compressing.  In numerical evolutions, the appearance of a spike in the metric field $A$, which is known as grid or slice stretching \cite{AlcubierreBook}, is a common indicator for the formation of a black hole.  Lastly, we show in Fig.\ \ref{fig:alpha} that the lapse function, $\alpha$, collapses, which is another commonly used indicator for the formation of a black hole \cite{BaumgarteBook, AlcubierreBook}.

We have found substantial evidence that black holes form in the EDM system.  Further, we have found that any null geodesic that travels through the wormhole becomes trapped inside a black hole and is unable to travel arbitrarily far away on the opposite side.  We have performed many simulations using different asymmetric static EDM solutions as initial data.  For example, we have considered values of $\bar{\mu}$ and $\bar{e}$ that are smaller by up to a couple orders of magnitude and considered larger values of $f_0$ than presented.  We have also computed different geodesics, including geodesics that originate on the left and travel to the right through the wormhole.  In all cases, the results are qualitatively similar to what we have presented in this section.  Of course, we cannot rule out that there exists some static EDM solution which, when numerically evolved, behaves differently.  Nonetheless, our results lead us to conclude that EDM wormholes are not traversable.


\section{Conclusion}
\label{sec:conclusion}

EDM wormholes are asymptotically flat wormhole solutions in general relativity that do not make use of exotic matter.  They are formed from two charged spin-1/2 fermions, as described by the charged Dirac equation minimally coupled to general relativity.  The static asymmetric solutions are regular everywhere and violate the null energy condition, which suggests that they may be traversable.  To determine if in fact they are traversable, we constructed a time dependent EDM model.  We then used static asymmetric wormhole solutions as initial data and numerically evolved them forward in time.

In all simulations we performed, we found convincing numerical evidence that black holes form in our system which are connected by the wormhole.  We computed null geodesics in this geometry to see if a signal could travel through the wormhole and make it arbitrarily far away on the opposite side.  In all cases considered, geodesics that crossed the wormhole were trapped inside a black hole.  These results led us to conclude that EDM wormholes are not traversable.

An important takeaway is that violation of the null energy condition is an insufficient condition for determining if a static wormhole solution is traversable. Black holes may form so quickly that any signal traveling through the wormhole will be caught inside one.  To determine if a wormhole is traversable, it may be necessary to make a time dependent analysis.


\appendix

\section{Metric and vierbein}
\label{app:metric}

In this appendix, we present results for the general spherically symmetric metric and the corresponding vierbein.  We use the vierbein to couple spinors to curved space.


\subsection{Metric and field equations}

Our code for simulating wormholes is based on the standard $3+1$ foliation of spacetime \cite{AlcubierreBook, BaumgarteBook}, in which spacetime is foliated into a continuum of time slices, where each time slice is a spatial hypersurface.  The general spherically symmetric metric in this formalism can be written
\begin{equation} \label{general metric}
\begin{split}
ds^2 &= -\left(\alpha^2 - A \beta^{r2}\right) dt^2 + 2 A \beta^r dt dr
 + dl^2
\\
dl^2 &=  A dr^2 + C \left( d\theta^2 + \sin^2\theta d\phi^2 \right),
\end{split}
\end{equation}
where $\alpha(t,r)$, $\beta^r(t,r)$, $A(t,r)$, and $C(t,r)$ parametrize the metric and $dl^2$ is the spatial metric on a time slice.  In addition to these four metric fields, we have also the extrinsic curvature $K\indices{^i_j}(t,r)$, whose two independent and nonvanishing components are $K\indices{^r_r}$ and $K_T \equiv 2K\indices{^\theta_\theta} = 2 K\indices{^\phi_\phi}$.

The Einstein field equations, $G\indices{^\mu_\nu} = 8\pi G T\indices{^\mu_\nu}$, where for completeness we include the gravitational constant $G$, lead to the evolution equations \cite{kain}
\begin{widetext}
\begin{equation} \label{evo eqs} 
\begin{split}
\partial_t A &=   - 2\alpha A K\indices{^r_r}
+ \beta^r \partial_r A + 2 A \partial_r \beta^r
\\
\partial_t C &= -\alpha C  K_T + \beta^r \partial_r C
\\
\partial_t K\indices{^r_r} 
&=
\alpha \left[ \frac{(\partial_r C)^2}{4AC^2}
- \frac{1}{C} 
+ (K\indices{^r_r})^2 
- \frac{1}{4}K_T^2 
+ 4\pi G (S +  \rho - 2 S\indices{^r_r} )
\right]
- \frac{\partial_r^2 \alpha}{A} + \frac{(\partial_r \alpha)(\partial_r A)}{2A^2}
+ \beta^r \partial_r K\indices{^r_r}
\\
\partial_t K_T
&= \alpha \left[ \frac{1}{C} 
- \frac{(\partial_r C)^2 }{4AC^2}
+ \frac{3}{4} (K_T)^2 
+ 8\pi G S\indices{^r_r}
\right]
- \frac{(\partial_r \alpha)(\partial_r C)}{AC}
+ \beta^r \left[
\frac{\partial_r C}{2C} \left(2 K\indices{^r_r} -  K_T \right) 
-8\pi G S_r
\right]
\end{split}
\end{equation}
and to the Hamiltonian and momentum constraint equations,
\begin{equation} \label{con eqs}
\begin{split}
\partial_r^2 C &=
A + \frac{(\partial_r A)(\partial_r C)}{2A} 
+ \frac{(\partial_r C)^2}{4C} 
+ ACK_T \left( K\indices{^r_r} + \frac{1}{4} K_T \right)
- 8\pi G AC \rho
\\
\partial_r K_T &=  
\frac{\partial_r C}{2C} \left(2 K\indices{^r_r} -  K_T \right)  -8\pi G S_r.
\end{split}
\end{equation}
\end{widetext}
The matter functions in these equations are functions of the energy-momentum tensor and are given by

\begin{equation} \label{matter functions defs}
\begin{split}
\rho &= n^\mu n^\nu T_{\mu\nu}
\\
S\indices{^r_r} &= \gamma^{rr} T_{rr}
\\
S\indices{^\theta_\theta} &= \gamma^{\theta\theta} T_{\theta\theta}
\\
S_r &= - n^\mu T_{\mu r}
\end{split}
\end{equation}
and $S = S\indices{^r_r} + 2 S\indices{^\theta_\theta}$, where $\gamma_{ij} = \text{diag}(A,C, C\sin^2\theta)$ is the spatial metric, $\gamma^{ij}$ is its inverse, and
\begin{equation} \label{n up mu}
n^\mu = (1/\alpha, -\beta^r/\alpha,0,0)
\end{equation}
is the timelike unit vector normal to a time slice.


\subsection{Vierbein and spinor connection}

We use the vierbein formalism to couple spinors to curved space \cite{Weinberg:1972kfs, Carroll:2004st, Freedman:2012zz}.  The vierbein, $e\indices{_a^\mu}$, is defined by
\begin{equation}
g_{\mu\nu} = e_{a\mu} e\indices{^a_\nu}, \qquad
\eta_{ab} = e_{a\mu} e\indices{_b^\mu},
\end{equation}
where $g_{\mu\nu}$ is the curved space metric and $\eta_{ab}$ is the flat space metric.  We use Greek letters for curved space indices and Latin letters from the beginning of the alphabet for flat space indices.  When specifying explicit components, we use $\mu = t,r,\theta,\phi$ and $a = 0,1,2,3$.

Our Lagrangian for fermions is given as the top equation in (\ref{Lagrangian 2}):
\begin{equation} \label{app Lagrangian}
\mathcal{L}_\psi^x = \frac{1}{2} \left[
\bar{\psi}_x \gamma^\mu D_\mu \psi_x
- (D_\mu \bar{\psi}_x) \gamma^\mu \psi_x
\right] - \mu \bar{\psi}_x \psi_x.
\end{equation}
The $\gamma^\mu$ are curved spaced $\gamma$-matrices, as indicated by their Greek index.  Flat space $\gamma$-matrices, $\gamma^a$, are related to curved spaced ones via the vierbein,
\begin{equation}
\gamma^\mu = e\indices{_a^\mu} \gamma^a, \qquad
\gamma^a = e\indices{^a_\mu} \gamma^\mu,
\end{equation}
where
\begin{equation}
\{\gamma^\mu, \gamma^\nu\} = 2 g^{\mu\nu}, \qquad
\{\gamma^a, \gamma^b\} = 2\eta^{ab}.
\end{equation}
Our convention for the adjoint spinor is
\begin{equation}
\bar{\psi}_x = \psi_x^\dag (-i \gamma^0).
\end{equation}
The covariant derivatives in (\ref{app Lagrangian}) are defined by
\begin{equation}
\begin{split}
D_\mu \psi^x &= \nabla_\mu \psi^x - ie \mathcal{A}_\mu \psi
\\
D_\mu \bar{\psi}^x &= \nabla_\mu \bar{\psi}^x + ie \mathcal{A}_\mu \bar{\psi}^x,
\end{split}
\end{equation}
where
\begin{equation}
\begin{split}
\nabla_\mu \psi_x &= \partial_\mu \psi_x + \frac{1}{4} \omega\indices{_\mu^{ab}} \gamma_{ab}  \psi_x
\\
\nabla_\mu \bar{\psi}_x &= \partial_\mu \bar{\psi}_x - \frac{1}{4} \bar{\psi}_x \omega\indices{_\mu^{ab}} \gamma_{ab},
\end{split}
\end{equation}
where $\gamma_{ab} \equiv \gamma_{[a}\gamma_{b]} = [\gamma_a,\gamma_b]/2$ and $\omega\indices{_\mu^{ab}}$ is the spin connection,
\begin{align}
w_{\mu ab} &= 
\frac{1}{2} e\indices{_a^\alpha} (\partial_\mu e_{b\alpha} - \partial_\alpha e_{b\mu} )
+ \frac{1}{2} e\indices{_b^\beta}
(\partial_\beta e_{a\mu} - \partial_\mu e_{a\beta} )
\notag
\\
&\qquad - \frac{1}{2} 
e\indices{^c_\mu}
e\indices{_a^\alpha}
e\indices{_b^\beta}  (\partial_\alpha e_{c\beta} - \partial_\beta e_{c\alpha} ).
\end{align}
Since we assume metric compatibility, $\nabla_\lambda g_{\mu\nu} = 0$, we have $\omega_{\mu ab} = -\omega_{\mu ba}$.

When deriving equations, it can help to choose a specific representation for the $\gamma$-matrices and vierbein.  Following \cite{Ventrella:2003fu, Daka:2019iix}, we use the Dirac representation for flat space $\gamma$-matrices,
\begin{equation} \label{Dirac rep}
\gamma^0 = i
\begin{pmatrix}
1 & 0 \\ 0 & -1
\end{pmatrix},
\qquad
\gamma^j = i
\begin{pmatrix}
0 & \sigma^j \\ -\sigma^j & 0
\end{pmatrix},
\end{equation}
where $j=1,2,3$ and where the $\sigma^j$ are the standard Pauli matrices,
\begin{equation}
\sigma^1 = 
\begin{pmatrix}
0 & 1 \\ 1 & 0
\end{pmatrix},
\quad
\sigma^2 = 
\begin{pmatrix}
0 & -i \\ i & 0
\end{pmatrix}, 
\quad
\sigma^3 = 
\begin{pmatrix}
1 & 0 \\ 0 & -1
\end{pmatrix}.
\end{equation}  
For the vierbein, we use
\begin{equation} \label{gamma gamma}
\begin{aligned} 
\gamma^t &= \frac{\gamma^0}{\alpha},&
\quad
\gamma^r &= \frac{\gamma^3}{\sqrt{A}} - \beta^r \frac{\gamma^0}{\alpha},
\\
\gamma^\theta &= \frac{\gamma^2}{\sqrt{C}}, &
\gamma^\phi &= \frac{\gamma^1}{\sqrt{C} \sin\theta},
\end{aligned}
\end{equation}
which also defines the curved spaced $\gamma$-matrices.  This choice for the vierbein associates the angular components, $\theta$ and $\phi$, with the off-diagonal Pauli matrices, $\sigma^1$ and $\sigma^2$.  This association helps with separating out the angular dependence, which we do in Appendix \ref{app:equations}.  We note that this vierbein reduces to the one used in \cite{Ventrella:2003fu, Daka:2019iix} for $\beta^r = 0$ and $C = r^2$.

We end this appendix with the components of the spinor connection, $\Gamma_\mu \equiv -\omega\indices{_\mu^{ab}} \gamma_{ab}/4$, for our choice of vierbein,
\begin{align}
\Gamma_t &= {\gamma}^0 {\gamma}^3 \frac{1}{4 \alpha \sqrt{A}} \left[ 2 \alpha \alpha' + \beta^r (\dot{A} - \beta^r A' - 2 A \beta^{r \prime} ) \right]
\notag
\\
\Gamma_r &= {\gamma}^0 {\gamma}^3 \frac{1}{4 \alpha \sqrt{A}} \left( \dot{A} - \beta^r A' - 2 A \beta^{r \prime} \right)
\notag
\\
\Gamma_\theta &= {\gamma}^3 {\gamma}^2 \frac{C'}{4 \sqrt{AC}} + {\gamma}^0 {\gamma}^2 \frac{1}{4 \alpha \sqrt{C}} \left( \dot{C} - \beta^r C' \right)
\notag
\\
\Gamma_\phi &= {\gamma}^3 {\gamma}^1 \frac{C' \sin\theta}{4 \sqrt{AC}}
+{\gamma}^2 {\gamma}^1 \frac{\cos\theta}{2}
\notag
\\
&\qquad
+ {\gamma}^0 {\gamma}^1 \frac{\sin\theta}{4\alpha \sqrt{C}} \left( \dot{C} - \beta^r C' \right),
\label{spinor connection}
\end{align}
where, in the appendices, a prime denotes an $r$-derivative and a dot denotes a $t$-derivative.


\section{Equations of motion and the energy-momentum tensor}
\label{app:equations}

In this appendix, we present the equations of motion and the energy-momentum tensor for both fermions and bosons for our time dependent EDM model.  For fermions, we give a detailed derivation of the Dirac spinor ansatz.


\subsection{Fermions}


\subsubsection{Equations of motion}

The equations of motion for fermions follow from the Lagrangian in (\ref{app Lagrangian}),
\begin{equation} \label{fermion eom}
\gamma^\mu (\partial_\mu - \Gamma_\mu - ie \mathcal{A}_\mu) \psi_x - \mu \psi_x= 0,
\end{equation}
where $\Gamma_\mu$ is the spinor connection given in (\ref{spinor connection}).  In a spherically symmetric system, $\psi_x = \psi_x(t,r,\theta,\phi)$, $\mathcal{A}_\mu = \mathcal{A}_\mu(t,r)$, and $\mathcal{A}_\mu = (\mathcal{A}_t, \mathcal{A}_r, 0,0)$.  To reduce clutter, we drop the subscripted $x$ in the following.

The standard approach to solving the equations of motion is to follow Unruh \cite{Unruh:1973bda} and Chandrasekhar \cite{Chandrasekhar:1976ap, Chandrasekhar:1985kt} and look for separable solutions
\begin{equation}
\psi = 
\begin{pmatrix}
\psi_1 (t,r,\theta,\phi) \\
\psi_2 (t,r,\theta,\phi) \\
\psi_3 (t,r,\theta,\phi) \\
\psi_4 (t,r,\theta,\phi)
\end{pmatrix}
=
\begin{pmatrix}
R_1(t,r) \Theta_1(\theta,\phi) \\
R_2(t,r) \Theta_2(\theta,\phi) \\
R_3(t,r) \Theta_3(\theta,\phi) \\
R_4(t,r) \Theta_4(\theta,\phi)
\end{pmatrix},
\end{equation}
where the $R$'s and $\Theta$'s are in general complex.  Plugging this into (\ref{fermion eom}), using the metric in (\ref{general metric}), the Dirac representation in (\ref{Dirac rep}), and the vierbein in (\ref{gamma gamma}), we find the four equations
\begin{widetext}
\begin{equation} \label{fermion eom 2}
\begin{split}
&\frac{i\sqrt{C}}{\alpha} \left(\frac{\dot{R}_1}{R_1} - \beta^r \frac{R_1'}{R_1} + W \right)
\left(+\frac{R_1}{R_4}\frac{\Theta_1}{\Theta_3}\right)
+ \frac{i\sqrt{C}}{\sqrt{A}} \left(\frac{R_3'}{R_3} + V \right) \left(+\frac{R_3}{R_4} \right)
- \mu \sqrt{C}\left(\frac{R_1}{R_4} \frac{\Theta_1}{\Theta_3}\right)
\\
&\qquad= - \left(\frac{\partial_\theta \Theta_4}{\Theta_4} + \frac{\cot\theta}{2} \right) \left(+\frac{\Theta_4}{\Theta3}\right)
- \frac{i}{\sin\theta} \frac{\partial_\phi \Theta_4}{\Theta_4} \left(+\frac{\Theta_4}{\Theta_3}\right)
\\
&\frac{i\sqrt{C}}{\alpha} \left(\frac{\dot{R}_2}{R_2} - \beta^r \frac{R_2'}{R_2} + W \right)
\left(+\frac{R_2}{R_3}\frac{\Theta_2}{\Theta_4}\right)
+ \frac{i\sqrt{C}}{\sqrt{A}} \left(\frac{R_4'}{R_4} + V \right) \left(-\frac{R_4}{R_3} \right)
- \mu \sqrt{C}\left(\frac{R_2}{R_3} \frac{\Theta_2}{\Theta_4}\right)
\\
&\qquad= - \left(\frac{\partial_\theta \Theta_3}{\Theta_3} + \frac{\cot\theta}{2} \right) \left(- \frac{\Theta_3}{\Theta_4}\right)
- \frac{i}{\sin\theta} \frac{\partial_\phi \Theta_3}{\Theta_3} \left(+\frac{\Theta_3}{\Theta_4}\right)
\\
&\frac{i\sqrt{C}}{\alpha} \left(\frac{\dot{R}_3}{R_3} - \beta^r \frac{R_3'}{R_3} + W \right)
\left(-\frac{R_3}{R_2}\frac{\Theta_3}{\Theta_1}\right)
+ \frac{i\sqrt{C}}{\sqrt{A}} \left(\frac{R_1'}{R_1} + V \right) \left(-\frac{R_1}{R_2} \right)
- \mu \sqrt{C} \left(\frac{R_3}{R_2} \frac{\Theta_3}{\Theta_1} \right)
\\
&\qquad= -  \left(\frac{\partial_\theta \Theta_2}{\Theta_2} + \frac{\cot\theta}{2} \right) \left(-\frac{\Theta_2}{\Theta_1}\right)
- \frac{i}{\sin\theta} \frac{\partial_\phi \Theta_2}{\Theta_2} \left(-\frac{\Theta_2}{\Theta_1}\right)
\\
&\frac{i\sqrt{C}}{\alpha} \left(\frac{\dot{R}_4}{R_4} - \beta^r \frac{R_4'}{R_4} + W \right)
\left(-\frac{R_4}{R_1}\frac{\Theta_4}{\Theta_2}\right)
+ \frac{i\sqrt{C}}{\sqrt{A}} \left(\frac{R_2'}{R_2} + V \right) \left(+\frac{R_2}{R_1} \right)
- \mu \sqrt{C} \left(\frac{R_4}{R_1} \frac{\Theta_4}{\Theta_2} \right)
\\
&\qquad= -  \left(\frac{\partial_\theta \Theta_1}{\Theta_1} + \frac{\cot\theta}{2} \right) \left(+ \frac{\Theta_1}{\Theta_2}\right)
- \frac{i}{\sin\theta} \frac{\partial_\phi \Theta_1}{\Theta_1} \left(- \frac{\Theta_1}{\Theta_2}\right),
\end{split}
\end{equation}
\end{widetext}
where
\begin{align} \label{W V def}
W &\equiv  - ie (\mathcal{A}_t - \beta^r \mathcal{A}_r)
+ \frac{\dot{A} - \beta^r A' - 2 A \beta^{r \prime}}{4 A} 
+ \frac{ \dot{C} - \beta^r C'}{2  C}
\notag
\\
V &\equiv - ie \mathcal{A}_r
+ \frac{\alpha'}{2\alpha} + \frac{C' }{2C}.
\end{align}
We now make some assumptions, which go beyond just separating the variables, that reduce these four equations down to two.  We assume
\begin{equation} \label{R theta assumptions}
R_2 = i R_1, \quad
R_4 = i R_3, \quad
\Theta_3 = \Theta_1, \quad
\Theta_4 = -\Theta_2.
\end{equation}

The resulting angular equations can be written in terms of spin-weighted spherical harmonics, ${_s Y_{j m}}$, with spin weight $s$,
\begin{equation} \label{sYlm}
\eth_\pm^{(s)}  ({_s Y_{j m}}) = \pm \sqrt{(j \mp s)(j \pm s + 1)} ( {_{s\pm 1} Y_{j m}}),
\end{equation} 
where
\begin{equation}
\eth_\pm^{(s)}
= \mp \frac{i}{\sin\theta} \partial_\phi
 - \partial_\theta
 \pm s \cot\theta
\end{equation}
are the raising and lowering operators.  Specifically, the angular equations can be written as
\begin{equation} \label{raising/lowering}
\eth_+^{(-1/2)} \Theta_2 = - n \Theta_1, \qquad
\eth_-^{(+1/2)} \Theta_1 = n \Theta_2,
\end{equation}
where $n$ is the separation constant.  Comparing this with (\ref{sYlm}), we find that
\begin{equation}
\Theta_1 = {_{+1/2}Y_{j m}},
\qquad
\Theta_2 = {_{-1/2}Y_{j m}}.
\end{equation}
The energy-momentum tensor we present below indicates that, to preserve spherical symmetry, we require two or more fermions.  We consider only two fermions.  Following \cite{Ventrella:2003fu, Daka:2019iix}, for one of them we use $\Theta_1 = {_{1/2}Y_{(1/2)(1/2)}}$ and $\Theta_2 = {_{-1/2}Y_{(1/2)(1/2)}}$ and for the other we use  $\Theta_1 = {_{1/2}Y_{(1/2)(-1/2)}}$ and $\Theta_2 = {_{-1/2}Y_{(1/2)(-1/2)}}$, where
\begin{equation} \label{spin-weighted}
\begin{split}
{_{\pm {1/2}}Y_{(1/2)(1/2)}} &= \frac{1}{2\sqrt{\pi}} e^{i\phi/2} y_\pm (\theta),
\\
{_{\pm {1/2}}Y_{(1/2)(-1/2)}} &= \pm \frac{1}{2\sqrt{\pi}} e^{-i\phi/2} y_\mp(\theta),
\end{split}
\end{equation}
with
\begin{equation}
y_\pm(\theta) \equiv \sqrt{\frac{1 \mp \cos\theta}{2}} = 
\begin{array}{c}
\sin(\theta/2) \\ \cos(\theta/2)
\end{array}.
\end{equation}
Our choice of spin-weighted spherical harmonics fixes the separation constant to
\begin{equation} \label{separation constant}
n = -1.
\end{equation}

We turn now to the radial equations.  Our assumptions in (\ref{R theta assumptions}) reduced the four equations in (\ref{fermion eom 2}) to two equations for $R_1$ and $R_3$.  It is convenient to trade $R_1$ and $R_3$ for $F$ and $G$, defined by
\begin{equation} \label{R1 R2}
\begin{split}
R_1(t,r) &\equiv \frac{F(t,r)}{C^{1/2}(t,r) A^{1/4}(t,r)}
\\
R_3(t,r) &\equiv \frac{G(t,r)}{C^{1/2}(t,r) A^{1/4}(t,r)},
\end{split}
\end{equation}
so as to remove an inconvenient time derivative.  The radial equations of motion are
\begin{widetext}
\begin{equation} \label{complex F G eom}
\begin{split}
\partial_t F &= F \left(\beta^r \frac{F'}{F} + ie \mathcal{A}_t + \frac{  \beta^{r \prime}}{2} - ie \beta^r \mathcal{A}_r\right)
- \frac{\alpha G}{\sqrt{C}}
-  \frac{\alpha G}{\sqrt{A}} \left(\frac{G'}{G} 
- ie \mathcal{A}_r
+ \frac{\alpha'}{2\alpha} - \frac{A'}{4A}
 \right)
- i \mu \alpha F
\\
\partial_t G &= G \left(\beta^r \frac{G'}{G} + ie \mathcal{A}_t + \frac{ \beta^{r \prime}}{2} - ie \beta^r \mathcal{A}_r \right)
+ \frac{\alpha F}{\sqrt{C} }
-  \frac{\alpha F}{\sqrt{A} } \left(\frac{F'}{F} 
- ie \mathcal{A}_r
+ \frac{\alpha'}{2\alpha} - \frac{A'}{4A}
 \right)
+ i \mu \alpha G.
\end{split}
\end{equation}

Before separating the equations of motion into their real and imaginary parts, we note that the final form for the Dirac spinor is the Dirac spinor ansatz given in Eq.\ (\ref{Dirac ansatz}), which captures our assumptions in  (\ref{R theta assumptions}) and our choice of spin-weighted spherical harmonics in (\ref{spin-weighted}).  The fermion sector therefore depends on the two complex functions $F$ and $G$.  Separating these into real and imaginary parts,
\begin{equation} \label{real fields}
\begin{split}
F(t,r) &= F_1(t,r) + i F_2(t,r), 
\\
G(t,r) &= G_1(t,r) + i G_2(t,r),
\end{split}
\end{equation}
the equations of motion are
\begin{equation} \label{F G eom}
\begin{split}
\partial_t F_1 &= \left(\beta^r F'_1 + \frac{  \beta^{r \prime}}{2} F_1 \right)
- \frac{\alpha G_1}{\sqrt{C}}
-  \frac{\alpha} {\sqrt{A}} \left(G'_1
+ \frac{\alpha'}{2\alpha} G_1 - \frac{A'}{4A} G_1
 \right)
- \left[ F_2\left( e \mathcal{A}_t - e \beta^r \mathcal{A}_r
- \mu \alpha\right) 
+G_2\frac{\alpha}{\sqrt{A}} e \mathcal{A}_r\right]
\\
\partial_t F_2 &= \left(\beta^r F'_2 + \frac{  \beta^{r \prime}}{2} F_2 \right)
- \frac{\alpha G_2}{\sqrt{C}}
-  \frac{\alpha} {\sqrt{A}} \left(G'_2
+ \frac{\alpha'}{2\alpha} G_2 - \frac{A'}{4A} G_2
 \right)
+ \left[ F_1\left( e \mathcal{A}_t - e \beta^r \mathcal{A}_r
- \mu \alpha\right) 
+G_1\frac{\alpha}{\sqrt{A}} e \mathcal{A}_r\right]
\\
\partial_t G_1 &= \left(\beta^r G'_1  + \frac{ \beta^{r \prime}}{2} G_1 \right)
+ \frac{\alpha F_1}{\sqrt{C} }
-  \frac{\alpha }{\sqrt{A} } \left(F'_1
+ \frac{\alpha'}{2\alpha}F_1 - \frac{A'}{4A} F_1
 \right)
- \left[G_2\left( e \mathcal{A}_t - e \beta^r \mathcal{A}_r 
+ \mu \alpha \right)
+ F_2  \frac{\alpha}{\sqrt{A}}  e \mathcal{A}_r
\right]
\\
\partial_t G_2 &= \left(\beta^r G'_2  + \frac{ \beta^{r \prime}}{2} G_2 \right)
+ \frac{\alpha F_2}{\sqrt{C} }
-  \frac{\alpha }{\sqrt{A} } \left(F'_2
+ \frac{\alpha'}{2\alpha}F_2 - \frac{A'}{4A} F_2
 \right)
+ \left[G_1\left( e \mathcal{A}_t - e \beta^r \mathcal{A}_r 
+ \mu \alpha \right)
+ F_1  \frac{\alpha}{\sqrt{A}}  e \mathcal{A}_r
\right].
\end{split}
\end{equation}


\subsubsection{Energy-momentum tensor}

The energy-momentum tensor in the fermion sector is given by
\begin{equation}
\begin{split}
(T_\psi^x)_{\mu\nu} &= 
-\frac{1}{4} \left[
\bar{\psi}_x \gamma_\mu D_\nu \psi_x
+ \bar{\psi}_x \gamma_\nu D_\mu \psi_x
- (D_\mu \bar{\psi}_x) \gamma_\nu \psi_x
- (D_\nu \bar{\psi}_x) \gamma_\mu \psi_x
\right].
\end{split}
\end{equation}
For a system to be spherically symmetric, only the diagonal components and $T_{tr} = T_{rt}$ can be nonvanishing.  The Dirac spinor ansatz in (\ref{Dirac ansatz}) leads to nonvanishing $(T_\psi^\pm)_{t\phi}$ and $(T_\psi^\pm)_{r\phi}$, which breaks spherical symmetry.  However, 
\begin{equation}
(T_\psi)_{\mu\nu} = (T_\psi^+)_{\mu\nu} + (T_\psi^-)_{\mu\nu}
\end{equation}
is spherically symmetric, which explains why two (or more) fermions are necessary to preserve spherical symmetry.  The nonvanishing components of the spherically symmetric energy-momentum tensor are
\begin{align} 
(T_\psi)_{tt} 
&= - \frac{\alpha}{2\pi \sqrt{A} C} \text{Im} \left(F^* \dot{F} + G^* \dot{G} \right)
+ \frac{\beta^r}{2\pi C} \text{Im}\left(\dot{F} G^* + F^* \dot{G}\right)
+ \frac{e \mathcal{A}_t}{2\pi \sqrt{A} C} \left[ \alpha\left( |F|^2 + |G|^2 \right)
- 2\sqrt{A} \beta^r \text{Re} \left( F G^* \right) \right]
\notag
\\
(T_\psi)_{rr}
&= \frac{1}{2\pi C} \text{Im} \left(F' G^* + F^*G' \right)
- \frac{e \mathcal{A}_r}{\pi C} \text{Re} \left( F G^*\right)
\notag
\\
(T_\psi)_{\theta\theta} &= \frac{1}{2\pi \sqrt{AC}} \text{Im}(F^* G)
\notag
\\
(T_\psi)_{\phi\phi}
&=  (T_\psi)_{\theta\theta} \sin^2\theta
\notag
\\
(T_\psi)_{tr}
&= -\frac{\alpha}{4\pi \sqrt{A} C} \text{Im} \left(F^* F' + G^* G' \right)
+ \frac{\beta^r }{4\pi C} \text{Im} \left(F' G^* + F^*G' \right)
+ \frac{1}{4\pi C} \text{Im} \left(\dot{F} G^* + F^* \dot{G}\right)
\notag
\\
&\qquad
- \frac{e}{4\pi \sqrt{A} C} \left[  2\sqrt{A}\left( \mathcal{A}_t + \beta^r \mathcal{A}_r\right) \text{Re}\left(FG^* \right)
-\alpha \mathcal{A}_r\left(|F|^2 + |G|^2\right)
\right].
\label{fermion em tensor}
\end{align}

We gave the equations for the metric fields and the extrinsic curvature in (\ref{evo eqs}) and (\ref{con eqs}).  These equations depend upon the matter functions defined in (\ref{matter functions defs}), which are functions of the energy-momentum tensor.  Moving to the real fields defined in (\ref{real fields}), and using the energy-momentum tensor in (\ref{fermion em tensor}), we find
\begin{equation} \label{fermion matter functions}
\begin{split}
\rho_\psi &= 
\frac{1}{2\pi \sqrt{A} C} 
\biggl[
\frac{2}{\sqrt{C}} \left( F_1 G_2 - F_2 G_1 \right)
- \frac{2 e \mathcal{A}_r }{\sqrt{A}}\left( F_1 G_1 + F_2G_2 \right)
+\frac{1}{\sqrt{A}} \left( F_1 G_2' - F_1' G_2 - F_2 G_1' + F_2' G_1 \right)
\\
&\qquad\qquad\qquad
+ \mu (F_1^2 + F_2^2 - G_1^2 - G_2^2)
\biggr]
\\
(S_\psi)\indices{^r_r} 
&= \frac{1}{2\pi A C} \left[ 
F_1 G_2' - F_2 G_1' + F_2' G_1 - F_1' G_2
- 2 e \mathcal{A}_r \left( F_1 G_1 + F_2 G_2 \right) \right]
\\
(S_\psi)\indices{^\theta_\theta} 
&=\frac{F_1 G_2 - F_2 G_1}{2\pi A^{1/2} C^{3/2}}
\\
(S_\psi)_r 
&= \frac{F_1 {F}_2' - F_2 {F}_1' + G_1 {G}_2' - G_2 {G}_1'
- e  \mathcal{A}_r (F_1^2 + F_2^2 + G_1^2 + G_2^2) }{2\pi \sqrt{A} C} ,
\end{split}
\end{equation}
\end{widetext}
along with $S_\psi = (S_\psi)\indices{^r_r}  + 2(S_\psi)\indices{^\theta_\theta}$, where we used the equations of motion in (\ref{F G eom}) to write them in this form.


\subsection{Bosons}

The equations of motion for the gauge boson are
\begin{equation} \label{gauge eom}
\nabla_\mu F^{\mu\nu} = J^\nu,
\end{equation}
where
\begin{equation} \label{conserved current}
J^\nu = i e \sum_x \bar{\psi}_x \gamma^\nu \psi_x
\end{equation}
is a conserved current. To facilitate solving this, we define the auxiliary field
\begin{equation}
Y(t,r) \equiv \frac{C(t,r)}{\alpha(t,r) \sqrt{A(t,r)}} \left[ \dot{\mathcal{A}}_r(t,r) - \mathcal{A}'_t (t,r) \right],
\end{equation}
so that the equations of motion can be written
\begin{equation} \label{Ydot Yprime}
\partial_t Y = -\alpha \sqrt{A} C J^r, \qquad
\partial_r Y = \alpha \sqrt{A} C J^t.
\end{equation}
Our focus is with two fermions that follow from the Dirac spinor ansatz in (\ref{Dirac ansatz}), for which
\begin{equation}
\begin{split}
J^t &= - \frac{e}{2\pi \alpha\sqrt{A} C} \left( F_1^2 + F_2^2 + G_1^2 + G_2^2\right)
\\
J^r &= - \frac{e}{2\pi A C}
\biggl[ 2\left(F_1 G_1 + F_2 G_2 \right) 
\\
&\qquad\qquad
- \frac{\beta^r \sqrt{A}}{\alpha} \left( F_1^2 + F_2^2 + G_1^2 + G_2^2 \right)
\biggr].
\end{split}
\end{equation}

The energy-momentum tensor for the gauge boson is given by
\begin{equation}
(T_\mathcal{A})_{\mu\nu} = g^{\alpha\beta} F_{\mu \alpha} F_{\nu \beta} - \frac{1}{4} g_{\mu\nu} F_{\alpha \beta}F^{\alpha\beta}.
\end{equation}
The components work out to be
\begin{equation} \label{gauge EM tensor}
\begin{split}
(T_\mathcal{A})_{tt} &= \frac{Y^2}{2C^2} \left(\alpha^2 - A \beta^{r2} \right)
\\
(T_\mathcal{A})_{rr} &= -\frac{A Y^2}{2C^2} 
\\
(T_\mathcal{A})_{\theta\theta} &= \frac{Y^2}{2C}
\\
(T_\mathcal{A})_{\phi\phi} &= (T_\mathcal{A})_{\theta\theta} \sin^2\theta
\\
(T_\mathcal{A})_{tr} &= -\beta^r \frac{A Y^2}{2C^2}
\end{split}
\end{equation}
and the matter functions defined in Eq.\ (\ref{matter functions defs}) are
\begin{equation} \label{boson matter functions}
\begin{split}
\rho_\mathcal{A} &=  +\frac{Y^2}{2 C^2}
\\
(S_\mathcal{A})\indices{^r_r} &= - \frac{Y^2}{2 C^2}
\\
(S_\mathcal{A})\indices{^\theta_\theta} &= +\frac{Y^2}{2C^2}
\\
(S_\mathcal{A})_r &= 0,
\end{split}
\end{equation}
along with $S_\mathcal{A} = (S_\mathcal{A})\indices{^r_r} + 2 (S_\mathcal{A})\indices{^\theta_\theta}$.

It will be convenient to make a $U(1)$ gauge choice.  For our numerical simulations, we use Lorenz gauge,
\begin{equation}
\nabla_\mu \mathcal{A}^\mu = 0.
\end{equation}
Defining
\begin{equation}
\Omega \equiv \frac{ \sqrt{A} C}{\alpha} \left(\mathcal{A}_t - \beta^r \mathcal{A}_r\right),
\end{equation}
where $\Omega(t,r)$ is an auxiliary field, the Lorenz gauge condition can be written
\begin{equation}
\partial_t \Omega = \partial_r \left( \frac{\alpha C}{\sqrt{A}} \mathcal{A}_r + \beta^r \Omega \right).
\end{equation}
From the definition of $\Omega$,
\begin{equation}
\mathcal{A}_t = \frac{\alpha}{\sqrt{A} C} \Omega + \beta^r \mathcal{A}_r.
\end{equation}


\section{Static equations}
\label{app:static}

Static wormhole solutions will be used as initial data for our simulations, in addition to being studied in their own right.  By static solutions, we mean that spacetime is time independent.  In this appendix, we present the equations we solve for static EDM wormholes.

To find static solutions, we drop the time dependence for metric fields and set $\beta^r = K\indices{^r_r} = K_T = 0$. Under these assumptions, our metric equations are \cite{kain}
\begin{equation}  \label{static eqs}
\begin{split}
\partial_r^2 C &=
A + \frac{(\partial_r A)(\partial_r C)}{2A} 
+ \frac{(\partial_r C)^2}{4C}
- 8\pi G AC \rho
\\
0
&= \frac{1}{C} - \frac{(\partial_r \alpha)(\partial_r C)}{\alpha AC}  
- \frac{(\partial_r C)^2 }{4AC^2} 
+ 8\pi  G S\indices{^r_r}
\\
\partial^2_r \alpha& = 
\left(
\frac{\partial_r A}{2A}
- \frac{\partial_r C}{ C}
\right) \partial_r \alpha
+ 4\pi G \alpha A(\rho + S).
\end{split}
\end{equation}
The first equation is the Hamiltonian constraint equation in (\ref{con eqs}), the second follows from the $K_T$ evolution equation in (\ref{evo eqs}), and the third follows from combining the $K\indices{^r_r}$ and $K_T$ evolution equations in (\ref{evo eqs}).

The energy-momentum tensor must be time independent and diagonal.  In the fermion sector, this can be achieved by assuming
\begin{equation} \label{f g}
F(t,r) = f(r) e^{-i\omega t}, \qquad
G(t,r) = i g(r) e^{-i\omega t},
\end{equation}
where $f(r)$ and $g(r)$ are real functions and $\omega$ is a real constant.  In the boson sector, we take $A_\mu$ to be time independent.  We choose to work in radial gauge, where
\begin{equation}
\mathcal{A}_r = 0.
\end{equation} 

Under these assumptions, in the fermion sector, the equations of motion in (\ref{F G eom}) become
\begin{equation} \label{static f g eom}
\begin{split}
f' &= 
- f \left( \frac{\alpha'}{2\alpha} - \frac{A'}{4A} - 
\sqrt{\frac{A}{C}}\right)
- g \sqrt{A} \left( \mu + \frac{u}{\alpha} \right)
\\
g' &= 
- g \left( \frac{\alpha'}{2\alpha} - \frac{A'}{4A} + 
\sqrt{\frac{A}{C}}\right)
- f \sqrt{A} \left( \mu - \frac{u}{\alpha} \right),
\end{split}
\end{equation}
where
\begin{equation}
u(r) \equiv e \mathcal{A}_t(r) + \omega,
\end{equation}
the energy-momentum tensor components in (\ref{fermion em tensor}) reduce to
\begin{equation} \label{static T psi}
\begin{split}
(T_\psi)_{tt} 
&=  \frac{\alpha u \left(f^2 + g^2\right)}{2\pi \sqrt{A} C}
\\
(T_\psi)_{rr}
&= \frac{f g' - f' g}{2\pi C}
\\
(T_\psi)_{\theta\theta} 
&= \frac{f g}{2\pi \sqrt{AC}} 
\\
(T_\psi)_{\phi\phi}
&=  (T_\psi)_{\theta\theta}  \sin^2\theta,
\end{split}
\end{equation}
along with $(T_\psi)_{tr} = 0$, and the matter functions in (\ref{fermion matter functions}) become
\begin{equation}
\begin{split}
\rho_\psi &= \frac{u(f^2 + g^2)}{2\pi \alpha \sqrt{A} C}
\\
(S_\psi)\indices{^r_r} &=   \frac{f g' - f' g}{2\pi A C} 
\\
(S_\psi)\indices{^\theta_\theta} &=   \frac{f g}{2\pi A^{1/2}C^{3/2}},
\end{split}
\end{equation}
along with $(S_\psi)_r$ = 0.  For convenience, we give
\begin{equation}
f g' - f' g = 
\frac{u\sqrt{A}}{\alpha}(f^2 + g^2)
-  \mu \sqrt{A}(f^2 - g^2)
-2fg \sqrt{\frac{A}{C}},
\end{equation}
which is used in $(S_\psi)\indices{^r_r}$ and which follows directly from (\ref{static f g eom}).

In the boson sector, the second equation of motion in (\ref{Ydot Yprime}) becomes
\begin{equation}
u'' = -u' \left(\frac{C'}{C} - \frac{\alpha'}{\alpha} - \frac{A'}{2A} \right)
+ e^2 \frac{\alpha \sqrt{A}}{2\pi C} ( f^2 + g^2),
\end{equation}
the energy-momentum tensor components in (\ref{gauge EM tensor}) become
\begin{equation} \label{static T A}
\begin{split}
(T_\mathcal{A})_{tt}  &=\frac{u^{\prime 2}}{2e^2A} 
\\
(T_\mathcal{A})_{rr} &= -\frac{u^{\prime 2}}{2e^2\alpha^2} 
\\
(T_\mathcal{A})_{\theta\theta} &= \frac{C u^{\prime 2}}{2 e^2 \alpha^2 A}
\\
(T_\mathcal{A})_{\phi\phi} &= (T_\mathcal{A})_{\theta\theta} \sin^2\theta,
\end{split}
\end{equation}
along with $(T_\mathcal{A})_{tr} = 0$, and the matter functions in (\ref{boson matter functions}) become
\begin{equation}
\begin{split}
\rho_\mathcal{A} = - (S_\mathcal{A})\indices{^r_r} = (S_\mathcal{A})\indices{^\theta_\theta} =  \frac{u^{\prime 2}}{2e^2\alpha^2 A},
\end{split}
\end{equation}
along with $(S_\mathcal{A})_r = 0$.


\section{Relation of static equations to \cite{Konoplya:2021hsm}}
\label{app:comparison}

In Appendix \ref{app:static}, we gave the complete set of equations we use to solve for static EDM wormholes.  Our equations do not look the same as those used by Konoplya and Zhidenko (KZ) in \cite{Konoplya:2021hsm}, but, as we explain here, they are equivalent.  KZ write in \cite{Konoplya:2021hsm} that they use the Dirac spinor ansatz given by Herdeiro, Pombo, and Radu (HPR) in \cite{Herdeiro:2017fhv}, who describe their static fermions with the functions $f_\text{HPR}(r)$ and $g_\text{HPR}(r)$.  The relationships between our fields $f(r)$ and $g(r)$ and HPR's are \cite{Daka:2019iix}
\begin{equation}
f_\text{HPR} = \frac{g}{4 \sqrt{\pi} A^{1/4} C^{1/2} }, \qquad
g_\text{HPR} = \frac{f}{4 \sqrt{\pi} A^{1/4} C^{1/2} }.
\end{equation} 
The relationships between our quantities and KZ's are
\begin{equation}
\begin{aligned}
x &= r,& \qquad q &=e,
\\
N(x) &= \alpha(r),& \qquad \frac{[\partial_x r(x)]^2}{B^2(x)} &= A(r),
\\
r^2(x) &= C(r),& \qquad V(x) &= \mathcal{A}_t(r),
\\
F(x) &= f_\text{HPR}(r), &\qquad G(x) &= g_\text{HPR}(r),
\end{aligned}
\end{equation}
where KZ's quantities are on the left and ours are on the right.  We have the same definitions for $\mu$ and $\omega$.  Lastly, KZ uses units such that $G = 1/4\pi$, while we use $G=1$.  With these relations, our static equations are identical to theirs.


\section{Numerical methods}
\label{app:numerics}

We have developed a second order accurate code for simulating EDM wormholes.  Our code is based on the code used in \cite{kain}, to which we refer the reader for additional details.  

We use static solutions as initial data for our simulations.  In the matter sector, we have
\begin{align}
F_1(0,r) &= f(r),&
F_2(0,r) &= 0,
\notag\\
G_1(0,r) &= 0,&
G_2(0,r) &= g(r)
\notag\\
\mathcal{A}_t(0,r) &= \frac{u(r)}{e},&
\mathcal{A}_r(0,r) &= 0
\notag \\
Y(0,r) &= \frac{-C(r) u'(r)}{e \alpha(r) \sqrt{A(r)}},& 
\Omega(0,r) &= \frac{C(r) \sqrt{A(r)} u(r)}{e \alpha(r)},
\end{align}
where the static solutions are on the right-hand side.  Note that we are assuming a $U(1)$ gauge where $u(r) = e\mathcal{A}_t(r)$.  For metric fields, we use the static solutions for $A(0,r)$ and $C(0,r)$.  We have also $K\indices{^r_r}(0,r) = K_T(0,r) = 0$.

The metric field $\alpha$ is a gravitational gauge field, in that once initial data have been loaded onto the initial time slice, $\alpha$ can be chosen arbitrarily \cite{AlcubierreBook, BaumgarteBook}.  Wet set $\alpha(0,r) = 1$ and then evolve $\alpha$ using the harmonic slicing condition \cite{ BaumgarteBook},
\begin{equation}
\partial_t \alpha= -\alpha^2 (K\indices{^r_r} + K_T).
\end{equation}
Harmonic slicing is known to have decent singularly avoidance properties and we have found that it works well for the EDM system. $\beta^r$ is also a gravitational gauge field.  As mentioned in Sec.\ \ref{sec:EDM}, we set $\beta^r = 0$.

We solve evolution equations using the method of lines and third order Runge-Kutta.  We solve the momentum constraint using second order Runge-Kutta.  At the outer boundaries, we use outgoing one-dimensional wave equations for matter fields $F_1$, $F_2$, $G_1$, and $G_2$ \cite{Daka:2019iix} and outgoing spherical wave equations for all other fields whose evolution equations contain spatial derivatives.

We set the outer boundary at $\bar{r} = \pm100$ and have confirmed that reflections are negligible.  We use time step $\Delta \bar{t} = 0.5 \Delta \bar{r}$ and we use uniform grid spacing $\Delta \bar{r} = 0.0025$.

An apparent horizon satisfies \cite{AlcubierreBook, BaumgarteBook, Gonzalez:2008xk}
\begin{equation} \label{apparent horizon}
 \sqrt{A} C K_T \mp \partial_r C = 0,
\end{equation}
where we use the upper sign for $r > 0$ and the lower sign for $r < 0$.  For the general spherically symmetric metric in (\ref{general metric}), null geodesics, $r_\text{null}(t)$, are computed from
\begin{equation} \label{null def}
\frac{dr_\text{null}}{dt} = \pm \frac{\alpha}{\sqrt{A}} - \beta^r,
\end{equation}
where the upper sign is for a right-moving geodesic and the lower sign is for a left-moving geodesic.


\section{Code tests}
\label{app:code tests}

\begin{figure*}
\centering
\includegraphics[width=6.5in]{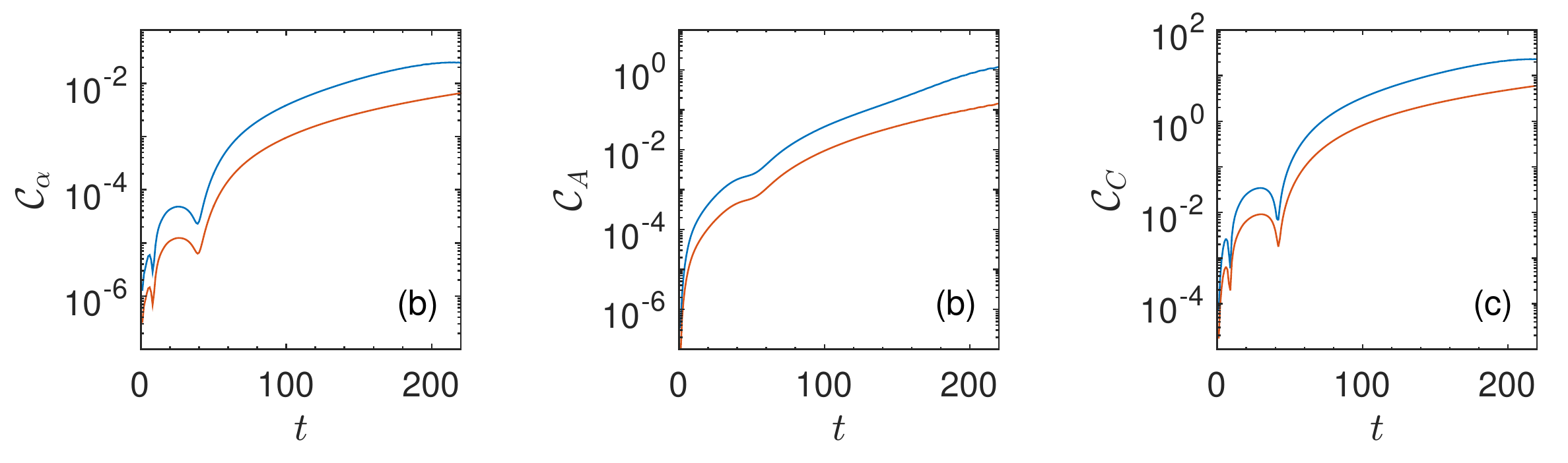}
\caption{The convergence function in (\ref{convergence function}) for  $\Delta r = 0.01$, 0.005, and 0.0025 is plotted for metric functions $\alpha$, $A$, and $C$ using the same initial data as used for the simulation shown in the top row of Fig.\ \ref{fig:R fsqgsq}.  The bottom curve drops by a factor of four compared to the top curve, indicating second order convergence.}
\label{fig:code tests}
\end{figure*} 

To test our code and to determine its level of convergence, we make use of the convergence function \cite{AlcubierreBook}
\begin{equation} \label{convergence function}
\mathcal{C}_f^{\Delta r_1, \Delta r_2} \equiv \| f^{\Delta r_1} - f^{\Delta r_2} \|,
\end{equation}
where $f^{\Delta r_1}$ is the value of an arbitrary field obtained numerically from an evolution equation using grid spacing $\Delta r_1$ and $\|$ indicates the $L^2$ norm across the computational grid.  Using grid spacings $\Delta r_1$, $\Delta r_2$, and $\Delta r_3$, where $\Delta r_1/\Delta r_2 = \Delta r_2/\Delta r_3 = 2$, an indication of second order convergence is that $\mathcal{C}_f^{\Delta r_1, \Delta r_2} / \mathcal{C}_f^{\Delta r_2, \Delta r_3} = 4$ \cite{AlcubierreBook}.

In Fig.\ \ref{fig:code tests}, we show results for the metric fields using $\Delta r = 0.01$, 0.005, and 0.0025.  In all plots, the lower curve drops by a factor of 4, indicating second order convergence.  We find similar results for our other simulations.  Although we do not show it here, we have confirmed that the Hamiltonian constraint in Eq.\ (\ref{con eqs beta0}) is satisfied on each time slice, as required.




\begin{thebibliography}{40}%
\makeatletter
\providecommand \@ifxundefined [1]{%
 \@ifx{#1\undefined}
}%
\providecommand \@ifnum [1]{%
 \ifnum #1\expandafter \@firstoftwo
 \else \expandafter \@secondoftwo
 \fi
}%
\providecommand \@ifx [1]{%
 \ifx #1\expandafter \@firstoftwo
 \else \expandafter \@secondoftwo
 \fi
}%
\providecommand \natexlab [1]{#1}%
\providecommand \enquote  [1]{``#1''}%
\providecommand \bibnamefont  [1]{#1}%
\providecommand \bibfnamefont [1]{#1}%
\providecommand \citenamefont [1]{#1}%
\providecommand \href@noop [0]{\@secondoftwo}%
\providecommand \href [0]{\begingroup \@sanitize@url \@href}%
\providecommand \@href[1]{\@@startlink{#1}\@@href}%
\providecommand \@@href[1]{\endgroup#1\@@endlink}%
\providecommand \@sanitize@url [0]{\catcode `\\12\catcode `\$12\catcode
  `\&12\catcode `\#12\catcode `\^12\catcode `\_12\catcode `\%12\relax}%
\providecommand \@@startlink[1]{}%
\providecommand \@@endlink[0]{}%
\providecommand \url  [0]{\begingroup\@sanitize@url \@url }%
\providecommand \@url [1]{\endgroup\@href {#1}{\urlprefix }}%
\providecommand \urlprefix  [0]{URL }%
\providecommand \Eprint [0]{\href }%
\providecommand \doibase [0]{https://doi.org/}%
\providecommand \selectlanguage [0]{\@gobble}%
\providecommand \bibinfo  [0]{\@secondoftwo}%
\providecommand \bibfield  [0]{\@secondoftwo}%
\providecommand \translation [1]{[#1]}%
\providecommand \BibitemOpen [0]{}%
\providecommand \bibitemStop [0]{}%
\providecommand \bibitemNoStop [0]{.\EOS\space}%
\providecommand \EOS [0]{\spacefactor3000\relax}%
\providecommand \BibitemShut  [1]{\csname bibitem#1\endcsname}%
\let\auto@bib@innerbib\@empty
\bibitem [{\citenamefont {Visser}(1996)}]{VisserBook}%
  \BibitemOpen
  \bibfield  {author} {\bibinfo {author} {\bibfnamefont {M.}~\bibnamefont
  {Visser}},\ }\href@noop {} {\emph {\bibinfo {title} {{Lorentzian wormholes:
  From Einstein to Hawking}}}}\ (\bibinfo  {publisher} {Springer-Verlag},\
  \bibinfo {year} {1996})\BibitemShut {NoStop}%
\bibitem [{\citenamefont {Lobo}(2017)}]{LoboBook}%
  \BibitemOpen
  \bibinfo {editor} {\bibfnamefont {F.~S.~N.}\ \bibnamefont {Lobo}},\ ed.,\
  \href@noop {} {\emph {\bibinfo {title} {{Wormholes, Warp Drives and Energy
  Conditions}}}}\ (\bibinfo  {publisher} {Springer},\ \bibinfo {year}
  {2017})\BibitemShut {NoStop}%
\bibitem [{\citenamefont {Ellis}(1973)}]{Ellis:1973yv}%
  \BibitemOpen
  \bibfield  {author} {\bibinfo {author} {\bibfnamefont {H.~G.}\ \bibnamefont
  {Ellis}},\ }\bibfield  {title} {\bibinfo {title} {{Ether flow through a
  drainhole - a particle model in general relativity}},\ }\href
  {https://doi.org/10.1063/1.1666161} {\bibfield  {journal} {\bibinfo
  {journal} {J. Math. Phys.}\ }\textbf {\bibinfo {volume} {14}},\ \bibinfo
  {pages} {104} (\bibinfo {year} {1973})}\BibitemShut {NoStop}%
\bibitem [{\citenamefont {Bronnikov}(1973)}]{Bronnikov:1973fh}%
  \BibitemOpen
  \bibfield  {author} {\bibinfo {author} {\bibfnamefont {K.~A.}\ \bibnamefont
  {Bronnikov}},\ }\bibfield  {title} {\bibinfo {title} {{Scalar-tensor theory
  and scalar charge}},\ }\href@noop {} {\bibfield  {journal} {\bibinfo
  {journal} {Acta Phys. Polon. B}\ }\textbf {\bibinfo {volume} {4}},\ \bibinfo
  {pages} {251} (\bibinfo {year} {1973})}\BibitemShut {NoStop}%
\bibitem [{\citenamefont
  {Armend\'ariz-Pic\'on}(2002)}]{Armendariz-Picon:2002gjc}%
  \BibitemOpen
  \bibfield  {author} {\bibinfo {author} {\bibfnamefont {C.}~\bibnamefont
  {Armend\'ariz-Pic\'on}},\ }\bibfield  {title} {\bibinfo {title} {{On a class
  of stable, traversable Lorentzian wormholes in classical general
  relativity}},\ }\href {https://doi.org/10.1103/PhysRevD.65.104010} {\bibfield
   {journal} {\bibinfo  {journal} {Phys. Rev. D}\ }\textbf {\bibinfo {volume}
  {65}},\ \bibinfo {pages} {104010} (\bibinfo {year} {2002})},\ \Eprint
  {https://arxiv.org/abs/gr-qc/0201027} {arXiv:gr-qc/0201027} \BibitemShut
  {NoStop}%
\bibitem [{\citenamefont {Bronnikov}\ and\ \citenamefont
  {Kim}(2003)}]{Bronnikov:2002rn}%
  \BibitemOpen
  \bibfield  {author} {\bibinfo {author} {\bibfnamefont {K.~A.}\ \bibnamefont
  {Bronnikov}}\ and\ \bibinfo {author} {\bibfnamefont {S.-W.}\ \bibnamefont
  {Kim}},\ }\bibfield  {title} {\bibinfo {title} {{Possible wormholes in a
  brane world}},\ }\href {https://doi.org/10.1103/PhysRevD.67.064027}
  {\bibfield  {journal} {\bibinfo  {journal} {Phys. Rev. D}\ }\textbf {\bibinfo
  {volume} {67}},\ \bibinfo {pages} {064027} (\bibinfo {year} {2003})},\
  \Eprint {https://arxiv.org/abs/gr-qc/0212112} {arXiv:gr-qc/0212112}
  \BibitemShut {NoStop}%
\bibitem [{\citenamefont {McFadden}\ and\ \citenamefont
  {Turok}(2005)}]{McFadden:2004ni}%
  \BibitemOpen
  \bibfield  {author} {\bibinfo {author} {\bibfnamefont {P.~L.}\ \bibnamefont
  {McFadden}}\ and\ \bibinfo {author} {\bibfnamefont {N.~G.}\ \bibnamefont
  {Turok}},\ }\bibfield  {title} {\bibinfo {title} {{Effective theory approach
  to brane world black holes}},\ }\href
  {https://doi.org/10.1103/PhysRevD.71.086004} {\bibfield  {journal} {\bibinfo
  {journal} {Phys. Rev. D}\ }\textbf {\bibinfo {volume} {71}},\ \bibinfo
  {pages} {086004} (\bibinfo {year} {2005})},\ \Eprint
  {https://arxiv.org/abs/hep-th/0412109} {arXiv:hep-th/0412109} \BibitemShut
  {NoStop}%
\bibitem [{\citenamefont {Kanti}\ \emph {et~al.}(2011)\citenamefont {Kanti},
  \citenamefont {Kleihaus},\ and\ \citenamefont {Kunz}}]{Kanti:2011jz}%
  \BibitemOpen
  \bibfield  {author} {\bibinfo {author} {\bibfnamefont {P.}~\bibnamefont
  {Kanti}}, \bibinfo {author} {\bibfnamefont {B.}~\bibnamefont {Kleihaus}},\
  and\ \bibinfo {author} {\bibfnamefont {J.}~\bibnamefont {Kunz}},\ }\bibfield
  {title} {\bibinfo {title} {{Wormholes in Dilatonic Einstein-Gauss-Bonnet
  Theory}},\ }\href {https://doi.org/10.1103/PhysRevLett.107.271101} {\bibfield
   {journal} {\bibinfo  {journal} {Phys. Rev. Lett.}\ }\textbf {\bibinfo
  {volume} {107}},\ \bibinfo {pages} {271101} (\bibinfo {year} {2011})},\
  \Eprint {https://arxiv.org/abs/1108.3003} {arXiv:1108.3003 [gr-qc]}
  \BibitemShut {NoStop}%
\bibitem [{\citenamefont {Kanti}\ \emph {et~al.}(2012)\citenamefont {Kanti},
  \citenamefont {Kleihaus},\ and\ \citenamefont {Kunz}}]{Kanti:2011yv}%
  \BibitemOpen
  \bibfield  {author} {\bibinfo {author} {\bibfnamefont {P.}~\bibnamefont
  {Kanti}}, \bibinfo {author} {\bibfnamefont {B.}~\bibnamefont {Kleihaus}},\
  and\ \bibinfo {author} {\bibfnamefont {J.}~\bibnamefont {Kunz}},\ }\bibfield
  {title} {\bibinfo {title} {{Stable Lorentzian Wormholes in Dilatonic
  Einstein-Gauss-Bonnet Theory}},\ }\href
  {https://doi.org/10.1103/PhysRevD.85.044007} {\bibfield  {journal} {\bibinfo
  {journal} {Phys. Rev. D}\ }\textbf {\bibinfo {volume} {85}},\ \bibinfo
  {pages} {044007} (\bibinfo {year} {2012})},\ \Eprint
  {https://arxiv.org/abs/1111.4049} {arXiv:1111.4049 [hep-th]} \BibitemShut
  {NoStop}%
\bibitem [{\citenamefont {Gao}\ \emph {et~al.}(2017)\citenamefont {Gao},
  \citenamefont {Jafferis},\ and\ \citenamefont {Wall}}]{Gao:2016bin}%
  \BibitemOpen
  \bibfield  {author} {\bibinfo {author} {\bibfnamefont {P.}~\bibnamefont
  {Gao}}, \bibinfo {author} {\bibfnamefont {D.~L.}\ \bibnamefont {Jafferis}},\
  and\ \bibinfo {author} {\bibfnamefont {A.~C.}\ \bibnamefont {Wall}},\
  }\bibfield  {title} {\bibinfo {title} {{Traversable Wormholes via a Double
  Trace Deformation}},\ }\href {https://doi.org/10.1007/JHEP12(2017)151}
  {\bibfield  {journal} {\bibinfo  {journal} {JHEP}\ }\textbf {\bibinfo
  {volume} {12}},\ \bibinfo {pages} {151}},\ \Eprint
  {https://arxiv.org/abs/1608.05687} {arXiv:1608.05687 [hep-th]} \BibitemShut
  {NoStop}%
\bibitem [{\citenamefont {Maldacena}\ \emph {et~al.}(2018)\citenamefont
  {Maldacena}, \citenamefont {Milekhin},\ and\ \citenamefont
  {Popov}}]{Maldacena:2018gjk}%
  \BibitemOpen
  \bibfield  {author} {\bibinfo {author} {\bibfnamefont {J.}~\bibnamefont
  {Maldacena}}, \bibinfo {author} {\bibfnamefont {A.}~\bibnamefont
  {Milekhin}},\ and\ \bibinfo {author} {\bibfnamefont {F.}~\bibnamefont
  {Popov}},\ }\bibfield  {title} {\bibinfo {title} {{Traversable wormholes in
  four dimensions}},\ }\href@noop {} {\  (\bibinfo {year} {2018})},\ \Eprint
  {https://arxiv.org/abs/1807.04726} {arXiv:1807.04726 [hep-th]} \BibitemShut
  {NoStop}%
\bibitem [{\citenamefont {Maldacena}\ and\ \citenamefont
  {Milekhin}(2021)}]{Maldacena:2020sxe}%
  \BibitemOpen
  \bibfield  {author} {\bibinfo {author} {\bibfnamefont {J.}~\bibnamefont
  {Maldacena}}\ and\ \bibinfo {author} {\bibfnamefont {A.}~\bibnamefont
  {Milekhin}},\ }\bibfield  {title} {\bibinfo {title} {{Humanly traversable
  wormholes}},\ }\href {https://doi.org/10.1103/PhysRevD.103.066007} {\bibfield
   {journal} {\bibinfo  {journal} {Phys. Rev. D}\ }\textbf {\bibinfo {volume}
  {103}},\ \bibinfo {pages} {066007} (\bibinfo {year} {2021})},\ \Eprint
  {https://arxiv.org/abs/2008.06618} {arXiv:2008.06618 [hep-th]} \BibitemShut
  {NoStop}%
\bibitem [{\citenamefont {Bl\'azquez-Salcedo}\ \emph
  {et~al.}(2021)\citenamefont {Bl\'azquez-Salcedo}, \citenamefont {Knoll},\
  and\ \citenamefont {Radu}}]{Blazquez-Salcedo:2020czn}%
  \BibitemOpen
  \bibfield  {author} {\bibinfo {author} {\bibfnamefont {J.~L.}\ \bibnamefont
  {Bl\'azquez-Salcedo}}, \bibinfo {author} {\bibfnamefont {C.}~\bibnamefont
  {Knoll}},\ and\ \bibinfo {author} {\bibfnamefont {E.}~\bibnamefont {Radu}},\
  }\bibfield  {title} {\bibinfo {title} {{Traversable wormholes in
  Einstein-Dirac-Maxwell theory}},\ }\href
  {https://doi.org/10.1103/PhysRevLett.126.101102} {\bibfield  {journal}
  {\bibinfo  {journal} {Phys. Rev. Lett.}\ }\textbf {\bibinfo {volume} {126}},\
  \bibinfo {pages} {101102} (\bibinfo {year} {2021})},\ \Eprint
  {https://arxiv.org/abs/2010.07317} {arXiv:2010.07317 [gr-qc]} \BibitemShut
  {NoStop}%
\bibitem [{\citenamefont {Bl\'azquez-Salcedo}\ \emph
  {et~al.}(2022)\citenamefont {Bl\'azquez-Salcedo}, \citenamefont {Knoll},\
  and\ \citenamefont {Radu}}]{Blazquez-Salcedo:2021udn}%
  \BibitemOpen
  \bibfield  {author} {\bibinfo {author} {\bibfnamefont {J.~L.}\ \bibnamefont
  {Bl\'azquez-Salcedo}}, \bibinfo {author} {\bibfnamefont {C.}~\bibnamefont
  {Knoll}},\ and\ \bibinfo {author} {\bibfnamefont {E.}~\bibnamefont {Radu}},\
  }\bibfield  {title} {\bibinfo {title}
  {{Einstein\textendash{}Dirac\textendash{}Maxwell wormholes: ansatz,
  construction and properties of symmetric solutions}},\ }\href
  {https://doi.org/10.1140/epjc/s10052-022-10488-6} {\bibfield  {journal}
  {\bibinfo  {journal} {Eur. Phys. J. C}\ }\textbf {\bibinfo {volume} {82}},\
  \bibinfo {pages} {533} (\bibinfo {year} {2022})},\ \Eprint
  {https://arxiv.org/abs/2108.12187} {arXiv:2108.12187 [gr-qc]} \BibitemShut
  {NoStop}%
\bibitem [{\citenamefont {Finster}\ \emph
  {et~al.}(1999{\natexlab{a}})\citenamefont {Finster}, \citenamefont
  {Smoller},\ and\ \citenamefont {Yau}}]{Finster:1998ws}%
  \BibitemOpen
  \bibfield  {author} {\bibinfo {author} {\bibfnamefont {F.}~\bibnamefont
  {Finster}}, \bibinfo {author} {\bibfnamefont {J.}~\bibnamefont {Smoller}},\
  and\ \bibinfo {author} {\bibfnamefont {S.-T.}\ \bibnamefont {Yau}},\
  }\bibfield  {title} {\bibinfo {title} {{Particle - like solutions of the
  Einstein-Dirac equations}},\ }\href
  {https://doi.org/10.1103/PhysRevD.59.104020} {\bibfield  {journal} {\bibinfo
  {journal} {Phys. Rev. D}\ }\textbf {\bibinfo {volume} {59}},\ \bibinfo
  {pages} {104020} (\bibinfo {year} {1999}{\natexlab{a}})},\ \Eprint
  {https://arxiv.org/abs/gr-qc/9801079} {arXiv:gr-qc/9801079 [gr-qc]}
  \BibitemShut {NoStop}%
\bibitem [{\citenamefont {Finster}\ \emph
  {et~al.}(1999{\natexlab{b}})\citenamefont {Finster}, \citenamefont
  {Smoller},\ and\ \citenamefont {Yau}}]{Finster:1998ux}%
  \BibitemOpen
  \bibfield  {author} {\bibinfo {author} {\bibfnamefont {F.}~\bibnamefont
  {Finster}}, \bibinfo {author} {\bibfnamefont {J.}~\bibnamefont {Smoller}},\
  and\ \bibinfo {author} {\bibfnamefont {S.-T.}\ \bibnamefont {Yau}},\
  }\bibfield  {title} {\bibinfo {title} {{Particle - like solutions of the
  Einstein-Dirac-Maxwell equations}},\ }\href
  {https://doi.org/10.1016/S0375-9601(99)00457-0} {\bibfield  {journal}
  {\bibinfo  {journal} {Phys. Lett. A}\ }\textbf {\bibinfo {volume} {259}},\
  \bibinfo {pages} {431} (\bibinfo {year} {1999}{\natexlab{b}})},\ \Eprint
  {https://arxiv.org/abs/gr-qc/9802012} {arXiv:gr-qc/9802012 [gr-qc]}
  \BibitemShut {NoStop}%
\bibitem [{\citenamefont {Finster}\ \emph
  {et~al.}(1999{\natexlab{c}})\citenamefont {Finster}, \citenamefont
  {Smoller},\ and\ \citenamefont {Yau}}]{Finster:1998ju}%
  \BibitemOpen
  \bibfield  {author} {\bibinfo {author} {\bibfnamefont {F.}~\bibnamefont
  {Finster}}, \bibinfo {author} {\bibfnamefont {J.}~\bibnamefont {Smoller}},\
  and\ \bibinfo {author} {\bibfnamefont {S.-T.}\ \bibnamefont {Yau}},\
  }\bibfield  {title} {\bibinfo {title} {{Nonexistence of black hole solutions
  for a spherically symmetric, static Einstein-Dirac-Maxwell system}},\ }\href
  {https://doi.org/10.1007/s002200050675} {\bibfield  {journal} {\bibinfo
  {journal} {Commun. Math. Phys.}\ }\textbf {\bibinfo {volume} {205}},\
  \bibinfo {pages} {249} (\bibinfo {year} {1999}{\natexlab{c}})},\ \Eprint
  {https://arxiv.org/abs/gr-qc/9810048} {arXiv:gr-qc/9810048} \BibitemShut
  {NoStop}%
\bibitem [{\citenamefont {Kain}(2023)}]{Kain:2023jgu}%
  \BibitemOpen
  \bibfield  {author} {\bibinfo {author} {\bibfnamefont {B.}~\bibnamefont
  {Kain}},\ }\bibfield  {title} {\bibinfo {title} {{Einstein-Dirac system in
  semiclassical gravity}},\ }\href
  {https://doi.org/10.1103/PhysRevD.107.124001} {\bibfield  {journal} {\bibinfo
   {journal} {Phys. Rev. D}\ }\textbf {\bibinfo {volume} {107}},\ \bibinfo
  {pages} {124001} (\bibinfo {year} {2023})},\ \Eprint
  {https://arxiv.org/abs/2304.10627} {arXiv:2304.10627 [gr-qc]} \BibitemShut
  {NoStop}%
\bibitem [{\citenamefont {Danielson}\ \emph {et~al.}(2021)\citenamefont
  {Danielson}, \citenamefont {Satishchandran}, \citenamefont {Wald},\ and\
  \citenamefont {Weinbaum}}]{Danielson:2021aor}%
  \BibitemOpen
  \bibfield  {author} {\bibinfo {author} {\bibfnamefont {D.~L.}\ \bibnamefont
  {Danielson}}, \bibinfo {author} {\bibfnamefont {G.}~\bibnamefont
  {Satishchandran}}, \bibinfo {author} {\bibfnamefont {R.~M.}\ \bibnamefont
  {Wald}},\ and\ \bibinfo {author} {\bibfnamefont {R.~J.}\ \bibnamefont
  {Weinbaum}},\ }\bibfield  {title} {\bibinfo {title}
  {{Bl\'azquez-Salcedo\textendash{}Knoll\textendash{}Radu wormholes are not
  solutions to the Einstein-Dirac-Maxwell equations}},\ }\href
  {https://doi.org/10.1103/PhysRevD.104.124055} {\bibfield  {journal} {\bibinfo
   {journal} {Phys. Rev. D}\ }\textbf {\bibinfo {volume} {104}},\ \bibinfo
  {pages} {124055} (\bibinfo {year} {2021})},\ \Eprint
  {https://arxiv.org/abs/2108.13361} {arXiv:2108.13361 [gr-qc]} \BibitemShut
  {NoStop}%
\bibitem [{\citenamefont {Konoplya}\ and\ \citenamefont
  {Zhidenko}(2022)}]{Konoplya:2021hsm}%
  \BibitemOpen
  \bibfield  {author} {\bibinfo {author} {\bibfnamefont {R.~A.}\ \bibnamefont
  {Konoplya}}\ and\ \bibinfo {author} {\bibfnamefont {A.}~\bibnamefont
  {Zhidenko}},\ }\bibfield  {title} {\bibinfo {title} {{Traversable Wormholes
  in General Relativity}},\ }\href
  {https://doi.org/10.1103/PhysRevLett.128.091104} {\bibfield  {journal}
  {\bibinfo  {journal} {Phys. Rev. Lett.}\ }\textbf {\bibinfo {volume} {128}},\
  \bibinfo {pages} {091104} (\bibinfo {year} {2022})},\ \Eprint
  {https://arxiv.org/abs/2106.05034} {arXiv:2106.05034 [gr-qc]} \BibitemShut
  {NoStop}%
\bibitem [{\citenamefont {Bolokhov}\ \emph {et~al.}(2021)\citenamefont
  {Bolokhov}, \citenamefont {Bronnikov}, \citenamefont {Krasnikov},\ and\
  \citenamefont {Skvortsova}}]{Bolokhov:2021fil}%
  \BibitemOpen
  \bibfield  {author} {\bibinfo {author} {\bibfnamefont {S.}~\bibnamefont
  {Bolokhov}}, \bibinfo {author} {\bibfnamefont {K.}~\bibnamefont {Bronnikov}},
  \bibinfo {author} {\bibfnamefont {S.}~\bibnamefont {Krasnikov}},\ and\
  \bibinfo {author} {\bibfnamefont {M.}~\bibnamefont {Skvortsova}},\ }\bibfield
   {title} {\bibinfo {title} {{A Note on \textquotedblleft{}Traversable
  Wormholes in Einstein\textendash{}Dirac\textendash{}Maxwell
  Theory\textquotedblright{}}},\ }\href
  {https://doi.org/10.1134/S0202289321040034} {\bibfield  {journal} {\bibinfo
  {journal} {Grav. Cosmol.}\ }\textbf {\bibinfo {volume} {27}},\ \bibinfo
  {pages} {401} (\bibinfo {year} {2021})},\ \Eprint
  {https://arxiv.org/abs/2104.10933} {arXiv:2104.10933 [gr-qc]} \BibitemShut
  {NoStop}%
\bibitem [{\citenamefont {Stuchl\'\i{}k}\ and\ \citenamefont
  {Vrba}(2021)}]{Stuchlik:2021guq}%
  \BibitemOpen
  \bibfield  {author} {\bibinfo {author} {\bibfnamefont {Z.}~\bibnamefont
  {Stuchl\'\i{}k}}\ and\ \bibinfo {author} {\bibfnamefont {J.}~\bibnamefont
  {Vrba}},\ }\bibfield  {title} {\bibinfo {title} {{Epicyclic orbits in the
  field of Einstein\textendash{}Dirac\textendash{}Maxwell traversable wormholes
  applied to the quasiperiodic oscillations observed in microquasars and active
  galactic nuclei}},\ }\href {https://doi.org/10.1140/epjp/s13360-021-02078-4}
  {\bibfield  {journal} {\bibinfo  {journal} {Eur. Phys. J. Plus}\ }\textbf
  {\bibinfo {volume} {136}},\ \bibinfo {pages} {1127} (\bibinfo {year}
  {2021})},\ \Eprint {https://arxiv.org/abs/2110.10569} {arXiv:2110.10569
  [gr-qc]} \BibitemShut {NoStop}%
\bibitem [{\citenamefont {Churilova}\ \emph {et~al.}(2021)\citenamefont
  {Churilova}, \citenamefont {Konoplya}, \citenamefont {Stuchlik},\ and\
  \citenamefont {Zhidenko}}]{Churilova:2021tgn}%
  \BibitemOpen
  \bibfield  {author} {\bibinfo {author} {\bibfnamefont {M.~S.}\ \bibnamefont
  {Churilova}}, \bibinfo {author} {\bibfnamefont {R.~A.}\ \bibnamefont
  {Konoplya}}, \bibinfo {author} {\bibfnamefont {Z.}~\bibnamefont {Stuchlik}},\
  and\ \bibinfo {author} {\bibfnamefont {A.}~\bibnamefont {Zhidenko}},\
  }\bibfield  {title} {\bibinfo {title} {{Wormholes without exotic matter:
  quasinormal modes, echoes and shadows}},\ }\href
  {https://doi.org/10.1088/1475-7516/2021/10/010} {\bibfield  {journal}
  {\bibinfo  {journal} {JCAP}\ }\textbf {\bibinfo {volume} {10}},\ \bibinfo
  {pages} {010}},\ \Eprint {https://arxiv.org/abs/2107.05977} {arXiv:2107.05977
  [gr-qc]} \BibitemShut {NoStop}%
\bibitem [{\citenamefont {Wang}\ \emph {et~al.}(2022)\citenamefont {Wang},
  \citenamefont {Wei},\ and\ \citenamefont {Liu}}]{Wang:2022aze}%
  \BibitemOpen
  \bibfield  {author} {\bibinfo {author} {\bibfnamefont {Y.-Q.}\ \bibnamefont
  {Wang}}, \bibinfo {author} {\bibfnamefont {S.-W.}\ \bibnamefont {Wei}},\ and\
  \bibinfo {author} {\bibfnamefont {Y.-X.}\ \bibnamefont {Liu}},\ }\bibfield
  {title} {\bibinfo {title} {{Comment on ``Traversable Wormholes in General
  Relativity''}},\ }\href@noop {} {\  (\bibinfo {year} {2022})},\ \Eprint
  {https://arxiv.org/abs/2206.12250} {arXiv:2206.12250 [gr-qc]} \BibitemShut
  {NoStop}%
\bibitem [{\citenamefont {Morris}\ and\ \citenamefont
  {Thorne}(1988)}]{Morris:1988cz}%
  \BibitemOpen
  \bibfield  {author} {\bibinfo {author} {\bibfnamefont {M.~S.}\ \bibnamefont
  {Morris}}\ and\ \bibinfo {author} {\bibfnamefont {K.~S.}\ \bibnamefont
  {Thorne}},\ }\bibfield  {title} {\bibinfo {title} {{Wormholes in space-time
  and their use for interstellar travel: A tool for teaching general
  relativity}},\ }\href {https://doi.org/10.1119/1.15620} {\bibfield  {journal}
  {\bibinfo  {journal} {Am. J. Phys.}\ }\textbf {\bibinfo {volume} {56}},\
  \bibinfo {pages} {395} (\bibinfo {year} {1988})}\BibitemShut {NoStop}%
\bibitem [{\citenamefont {Calhoun}\ \emph {et~al.}(2022)\citenamefont
  {Calhoun}, \citenamefont {Fay},\ and\ \citenamefont {Kain}}]{kain}%
  \BibitemOpen
  \bibfield  {author} {\bibinfo {author} {\bibfnamefont {K.}~\bibnamefont
  {Calhoun}}, \bibinfo {author} {\bibfnamefont {B.}~\bibnamefont {Fay}},\ and\
  \bibinfo {author} {\bibfnamefont {B.}~\bibnamefont {Kain}},\ }\bibfield
  {title} {\bibinfo {title} {{Matter traveling through a wormhole}},\ }\href
  {https://doi.org/10.1103/PhysRevD.106.104054} {\bibfield  {journal} {\bibinfo
   {journal} {Phys. Rev. D}\ }\textbf {\bibinfo {volume} {106}},\ \bibinfo
  {pages} {104054} (\bibinfo {year} {2022})},\ \Eprint
  {https://arxiv.org/abs/2210.04905} {arXiv:2210.04905 [gr-qc]} \BibitemShut
  {NoStop}%
\bibitem [{\citenamefont {Herdeiro}\ \emph {et~al.}(2017)\citenamefont
  {Herdeiro}, \citenamefont {Pombo},\ and\ \citenamefont
  {Radu}}]{Herdeiro:2017fhv}%
  \BibitemOpen
  \bibfield  {author} {\bibinfo {author} {\bibfnamefont {C.~A.~R.}\
  \bibnamefont {Herdeiro}}, \bibinfo {author} {\bibfnamefont {A.~M.}\
  \bibnamefont {Pombo}},\ and\ \bibinfo {author} {\bibfnamefont
  {E.}~\bibnamefont {Radu}},\ }\bibfield  {title} {\bibinfo {title}
  {{Asymptotically flat scalar, Dirac and Proca stars: discrete vs. continuous
  families of solutions}},\ }\href
  {https://doi.org/10.1016/j.physletb.2017.09.036} {\bibfield  {journal}
  {\bibinfo  {journal} {Phys. Lett. B}\ }\textbf {\bibinfo {volume} {773}},\
  \bibinfo {pages} {654} (\bibinfo {year} {2017})},\ \Eprint
  {https://arxiv.org/abs/1708.05674} {arXiv:1708.05674 [gr-qc]} \BibitemShut
  {NoStop}%
\bibitem [{\citenamefont {Baumgarte}\ and\ \citenamefont
  {Shapiro}(2010)}]{BaumgarteBook}%
  \BibitemOpen
  \bibfield  {author} {\bibinfo {author} {\bibfnamefont {T.~W.}\ \bibnamefont
  {Baumgarte}}\ and\ \bibinfo {author} {\bibfnamefont {S.~L.}\ \bibnamefont
  {Shapiro}},\ }\href@noop {} {\emph {\bibinfo {title} {{Numerical relativity:
  Solving Einstein's equations on the computer}}}}\ (\bibinfo  {publisher}
  {Cambridge},\ \bibinfo {year} {2010})\BibitemShut {NoStop}%
\bibitem [{\citenamefont {Anninos}\ \emph {et~al.}(1995)\citenamefont
  {Anninos}, \citenamefont {Bernstein}, \citenamefont {Brandt}, \citenamefont
  {Libson}, \citenamefont {Masso}, \citenamefont {Seidel}, \citenamefont
  {Smarr}, \citenamefont {Suen},\ and\ \citenamefont
  {Walker}}]{Anninos:1994ay}%
  \BibitemOpen
  \bibfield  {author} {\bibinfo {author} {\bibfnamefont {P.}~\bibnamefont
  {Anninos}}, \bibinfo {author} {\bibfnamefont {D.}~\bibnamefont {Bernstein}},
  \bibinfo {author} {\bibfnamefont {S.}~\bibnamefont {Brandt}}, \bibinfo
  {author} {\bibfnamefont {J.}~\bibnamefont {Libson}}, \bibinfo {author}
  {\bibfnamefont {J.}~\bibnamefont {Masso}}, \bibinfo {author} {\bibfnamefont
  {E.}~\bibnamefont {Seidel}}, \bibinfo {author} {\bibfnamefont
  {L.}~\bibnamefont {Smarr}}, \bibinfo {author} {\bibfnamefont {W.-M.}\
  \bibnamefont {Suen}},\ and\ \bibinfo {author} {\bibfnamefont
  {P.}~\bibnamefont {Walker}},\ }\bibfield  {title} {\bibinfo {title}
  {{Dynamics of apparent and event horizons}},\ }\href
  {https://doi.org/10.1103/PhysRevLett.74.630} {\bibfield  {journal} {\bibinfo
  {journal} {Phys. Rev. Lett.}\ }\textbf {\bibinfo {volume} {74}},\ \bibinfo
  {pages} {630} (\bibinfo {year} {1995})},\ \Eprint
  {https://arxiv.org/abs/gr-qc/9403011} {arXiv:gr-qc/9403011} \BibitemShut
  {NoStop}%
\bibitem [{\citenamefont {Libson}\ \emph {et~al.}(1996)\citenamefont {Libson},
  \citenamefont {Masso}, \citenamefont {Seidel}, \citenamefont {Suen},\ and\
  \citenamefont {Walker}}]{Libson:1994dk}%
  \BibitemOpen
  \bibfield  {author} {\bibinfo {author} {\bibfnamefont {J.}~\bibnamefont
  {Libson}}, \bibinfo {author} {\bibfnamefont {J.}~\bibnamefont {Masso}},
  \bibinfo {author} {\bibfnamefont {E.}~\bibnamefont {Seidel}}, \bibinfo
  {author} {\bibfnamefont {W.-M.}\ \bibnamefont {Suen}},\ and\ \bibinfo
  {author} {\bibfnamefont {P.}~\bibnamefont {Walker}},\ }\bibfield  {title}
  {\bibinfo {title} {{Event horizons in numerical relativity. 1: Methods and
  tests}},\ }\href {https://doi.org/10.1103/PhysRevD.53.4335} {\bibfield
  {journal} {\bibinfo  {journal} {Phys. Rev. D}\ }\textbf {\bibinfo {volume}
  {53}},\ \bibinfo {pages} {4335} (\bibinfo {year} {1996})},\ \Eprint
  {https://arxiv.org/abs/gr-qc/9412068} {arXiv:gr-qc/9412068} \BibitemShut
  {NoStop}%
\bibitem [{\citenamefont {Gonz\'alez}\ \emph {et~al.}(2009)\citenamefont
  {Gonz\'alez}, \citenamefont {Guzm\'an},\ and\ \citenamefont
  {Sarbach}}]{Gonzalez:2008xk}%
  \BibitemOpen
  \bibfield  {author} {\bibinfo {author} {\bibfnamefont {J.~A.}\ \bibnamefont
  {Gonz\'alez}}, \bibinfo {author} {\bibfnamefont {F.~S.}\ \bibnamefont
  {Guzm\'an}},\ and\ \bibinfo {author} {\bibfnamefont {O.}~\bibnamefont
  {Sarbach}},\ }\bibfield  {title} {\bibinfo {title} {{Instability of wormholes
  supported by a ghost scalar field. II. Nonlinear evolution}},\ }\href
  {https://doi.org/10.1088/0264-9381/26/1/015011} {\bibfield  {journal}
  {\bibinfo  {journal} {Class. Quant. Grav.}\ }\textbf {\bibinfo {volume}
  {26}},\ \bibinfo {pages} {015011} (\bibinfo {year} {2009})},\ \Eprint
  {https://arxiv.org/abs/0806.1370} {arXiv:0806.1370 [gr-qc]} \BibitemShut
  {NoStop}%
\bibitem [{\citenamefont {Alcubierre}(2008)}]{AlcubierreBook}%
  \BibitemOpen
  \bibfield  {author} {\bibinfo {author} {\bibfnamefont {M.}~\bibnamefont
  {Alcubierre}},\ }\href@noop {} {\emph {\bibinfo {title} {{Introduction to 3+1
  numerical relativity}}}}\ (\bibinfo  {publisher} {Oxford},\ \bibinfo {year}
  {2008})\BibitemShut {NoStop}%
\bibitem [{\citenamefont {Weinberg}(1972)}]{Weinberg:1972kfs}%
  \BibitemOpen
  \bibfield  {author} {\bibinfo {author} {\bibfnamefont {S.}~\bibnamefont
  {Weinberg}},\ }\href@noop {} {\emph {\bibinfo {title} {{Gravitation and
  Cosmology}}}}\ (\bibinfo  {publisher} {John Wiley and Sons},\ \bibinfo {year}
  {1972})\BibitemShut {NoStop}%
\bibitem [{\citenamefont {Carroll}(2004)}]{Carroll:2004st}%
  \BibitemOpen
  \bibfield  {author} {\bibinfo {author} {\bibfnamefont {S.~M.}\ \bibnamefont
  {Carroll}},\ }\href@noop {} {\emph {\bibinfo {title} {{Spacetime and
  geometry: An introduction to general relativity}}}}\ (\bibinfo  {publisher}
  {Addison-Wesley},\ \bibinfo {year} {2004})\BibitemShut {NoStop}%
\bibitem [{\citenamefont {Freedman}\ and\ \citenamefont
  {Van~Proeyen}(2012)}]{Freedman:2012zz}%
  \BibitemOpen
  \bibfield  {author} {\bibinfo {author} {\bibfnamefont {D.~Z.}\ \bibnamefont
  {Freedman}}\ and\ \bibinfo {author} {\bibfnamefont {A.}~\bibnamefont
  {Van~Proeyen}},\ }\href@noop {} {\emph {\bibinfo {title} {{Supergravity}}}}\
  (\bibinfo  {publisher} {Cambridge},\ \bibinfo {year} {2012})\BibitemShut
  {NoStop}%
\bibitem [{\citenamefont {Ventrella}\ and\ \citenamefont
  {Choptuik}(2003)}]{Ventrella:2003fu}%
  \BibitemOpen
  \bibfield  {author} {\bibinfo {author} {\bibfnamefont {J.~F.}\ \bibnamefont
  {Ventrella}}\ and\ \bibinfo {author} {\bibfnamefont {M.~W.}\ \bibnamefont
  {Choptuik}},\ }\bibfield  {title} {\bibinfo {title} {{Critical phenomena in
  the Einstein massless Dirac system}},\ }\href
  {https://doi.org/10.1103/PhysRevD.68.044020} {\bibfield  {journal} {\bibinfo
  {journal} {Phys. Rev. D}\ }\textbf {\bibinfo {volume} {68}},\ \bibinfo
  {pages} {044020} (\bibinfo {year} {2003})},\ \Eprint
  {https://arxiv.org/abs/gr-qc/0304007} {arXiv:gr-qc/0304007 [gr-qc]}
  \BibitemShut {NoStop}%
\bibitem [{\citenamefont {Daka}\ \emph {et~al.}(2019)\citenamefont {Daka},
  \citenamefont {Phan},\ and\ \citenamefont {Kain}}]{Daka:2019iix}%
  \BibitemOpen
  \bibfield  {author} {\bibinfo {author} {\bibfnamefont {E.}~\bibnamefont
  {Daka}}, \bibinfo {author} {\bibfnamefont {N.~N.}\ \bibnamefont {Phan}},\
  and\ \bibinfo {author} {\bibfnamefont {B.}~\bibnamefont {Kain}},\ }\bibfield
  {title} {\bibinfo {title} {{Perturbing the ground state of Dirac stars}},\
  }\href {https://doi.org/10.1103/PhysRevD.100.084042} {\bibfield  {journal}
  {\bibinfo  {journal} {Phys. Rev. D}\ }\textbf {\bibinfo {volume} {100}},\
  \bibinfo {pages} {084042} (\bibinfo {year} {2019})},\ \Eprint
  {https://arxiv.org/abs/1910.09415} {arXiv:1910.09415 [gr-qc]} \BibitemShut
  {NoStop}%
\bibitem [{\citenamefont {Unruh}(1973)}]{Unruh:1973bda}%
  \BibitemOpen
  \bibfield  {author} {\bibinfo {author} {\bibfnamefont {W.}~\bibnamefont
  {Unruh}},\ }\bibfield  {title} {\bibinfo {title} {{Separability of the
  Neutrino Equations in a Kerr Background}},\ }\href
  {https://doi.org/10.1103/PhysRevLett.31.1265} {\bibfield  {journal} {\bibinfo
   {journal} {Phys. Rev. Lett.}\ }\textbf {\bibinfo {volume} {31}},\ \bibinfo
  {pages} {1265} (\bibinfo {year} {1973})}\BibitemShut {NoStop}%
\bibitem [{\citenamefont {Chandrasekhar}(1976)}]{Chandrasekhar:1976ap}%
  \BibitemOpen
  \bibfield  {author} {\bibinfo {author} {\bibfnamefont {S.}~\bibnamefont
  {Chandrasekhar}},\ }\bibfield  {title} {\bibinfo {title} {{The Solution of
  Dirac's Equation in Kerr Geometry}},\ }\href
  {https://doi.org/10.1098/rspa.1976.0090} {\bibfield  {journal} {\bibinfo
  {journal} {Proc. Roy. Soc. Lond. A}\ }\textbf {\bibinfo {volume} {349}},\
  \bibinfo {pages} {571} (\bibinfo {year} {1976})}\BibitemShut {NoStop}%
\bibitem [{\citenamefont {Chandrasekhar}(1983)}]{Chandrasekhar:1985kt}%
  \BibitemOpen
  \bibfield  {author} {\bibinfo {author} {\bibfnamefont {S.}~\bibnamefont
  {Chandrasekhar}},\ }\href@noop {} {\emph {\bibinfo {title} {{The mathematical
  theory of black holes}}}}\ (\bibinfo  {publisher} {Oxford},\ \bibinfo {year}
  {1983})\BibitemShut {NoStop}%
\end{thebibliography}

%

\end{document}